\documentclass[twocolumn,showpacs,amsmath,amssymb,aps,prb]{revtex4-1}
\usepackage{graphicx}
\usepackage{dcolumn}
\usepackage{bm}
\usepackage[abs]{overpic}
\usepackage{color}

\begin{document}

\title{Manifestation of chirality in the vortex lattice in a two-dimensional topological superconductor}

\author{Evan D. B. Smith}
\affiliation{Department of Physics and Engineering Physics, University of Saskatchewan, 116 Science Place, Saskatoon, SK, S7N 5E2 Canada}

\author{K. Tanaka}
\email{Corresponding author: kat221@campus.usask.ca}
\affiliation{Department of Physics and Engineering Physics, University of Saskatchewan, 116 Science Place, Saskatoon, SK, S7N 5E2 Canada}

\author{Yuki Nagai}
\affiliation{CCSE, Japan  Atomic Energy Agency, 178-4-4, Wakashiba, Kashiwa, Chiba, 277-0871, Japan}

\date{\today}

\begin{abstract}
We study the vortex lattice in a two-dimensional $s$-wave topological superconductor with Rashba spin-orbit coupling and Zeeman field by solving the Bogoliubov-de Gennes equations self-consistently for the superconducting order parameter. We find that when spin-orbit coupling is relatively weak, one of the two underlying chiralities in the topological superconducting state can be strongly manifest in the vortex core structure and govern the response of the system to vorticity and a nonmagnetic impurity where the vortex is pinned. The Majorana zero mode in the vortex core is found to be robust against the nonmagnetic impurity in that it remains effectively a zero-energy bound state regardless of the impurity potential strength and the major chirality. The spin polarization of the Majorana bound state depends on the major chirality for weak spin-orbit coupling, while it is determined simply by the vorticity when spin-orbit coupling is relatively strong.
\end{abstract}

\pacs{74.20.-z, 74.20.Rp, 74.25.Uv, 75.70.Tj}
\maketitle

\section{\label{sec:introduction}Introduction}

One of the most promising models proposed so far for platforms to realise topological quantum computation is the two-dimensional (2D) $s$-wave topological superconductivity (TSC) model with Rashba spin-orbit (SO) coupling and Zeeman field.\cite{Sato2009,Sato2010,Sau2010,Alicea2010,Alicea2012} Sato, Takahashi, and Fujimoto have proposed the tight-binding model\cite{Sato2009,Sato2010} that can describe an $s$-wave superfluid of ultracold fermionic atoms in an optical lattice, where an effective SO interaction can be generated by spatially varying laser fields, or 2D $s$-wave TSC in a solid device. Such TSC can be realised, as proposed by Sau {\it et al.} in terms of the continuum model,\cite{Sau2010} in a semiconductor heterostructure, where a semiconductor with Rashba SO coupling is sandwiched between a conventional $s$-wave superconductor and a ferromagnetic insulator. While $s$-wave superconductivity is induced by the proximity effect, the ferromagnetic insulator can generate Zeeman coupling via exchange interactions, thus affecting only the spin degree of freedom in the semiconductor. A vortex in the model hosts a zero-energy Majorana bound state\cite{Sau2010,Tewari2010} and hence vortices in a 2D $s$-wave topological superconductor obey non-Abelian exchange statistics, like those in chiral $p$-wave superconductors.\cite{Kopnin1991,Read2000,Ivanov2001,Stern2004,Stone2006}

Inherent in the 2D $s$-wave TSC model are the two chiralities, $\sim{\rm sin}\,k_x \pm i\,{\rm sin}\,k_y$ in the tight-binding model\cite{Fujimoto2008,Sato2009p-wave} or $k_x\pm i k_y$ in the continuum model\cite{Zhang2008,Alicea2010,Shitade2015} (also for $|\bm{k}|\ll 1$ in the tight-binding model), where $\bm{k}=(k_x,k_y)$ is the wavevector. The noninteracting Fermi surface is split sideways by the Rashba SO interaction, which causes winding of spin as one goes around each Fermi surface,\cite{Frigeri2006} and the Zeeman field perpendicular to the 2D system favours one spin component along its direction over the other, resulting in two separate bands. With large enough Zeeman splitting and the chemical potential in the gap between the two bands, the system has a single Fermi surface on which spin is fixed for each $\bm{k}$ and thus becomes effectively spinless -- condition required for TSC that can support Majorana zero modes.\cite{Alicea2012} 
Although the Fermi surface has a certain chirality (spin winding), in principle the two intrinsic chiralities are always present in a 2D $s$-wave TSC state.\cite{Shitade2015}

It is known that in a spin-triplet chiral $p$-wave superconductor, vortices in the $p_x+ip_y$ and $p_x-ip_y$ states, which are degenerate at zero field, are not equivalent.\cite{Takigawa2001,Ichioka2002} The free energy is lower and the upper critical field is higher when the vorticity, i.e., the angular momentum carried by the supercurrent, is antiparallel to the chirality, i.e., the angular momentum of Cooper pairs in the condensate.\cite{Ichioka2002} Moreover, bound states in such an antiparallel vortex are more robust against nonmagnetic impurities due to cancellation of angular momenta between the supercurrent and the condensate; which makes the vortex core region $s$-wave-like and allows the Anderson theorem\cite{Anderson1959} to take effect.\cite{Kato2000,Kato2002} Most notably, in each of the chirality domains, the order parameter of the other chirality is induced around the vortex centre, more prominently for lower applied field.\cite{Takigawa2001,Ichioka2002}

The effects of nonmagnetic impurities on vortex bound states in a 2D $s$-wave topological superconductor, depending on the major chirality with respect to the vorticity, have been studied recently for a single vortex in the continuum model\cite{Sau2010} by means of a Green-function technique for calculating the impurity self-energy.\cite{Masaki2014,Masaki2015}  It has been found that when the major chirality is opposite to the vorticity, vortex bound states are less influenced by nonmagnetic impurities for relatively weak SO coupling, compared to the case where the major chirality is in the same direction as the vorticity. As to spinless $p_x\pm ip_y$ superconducting states per se, Ivanov has stated that though with slightly different structure of the quasiparticle eigenfunctions, the two types of vortices have the same low-energy spectra and braiding statistics.\cite{Ivanov2001} 

Volovik has studied the effects of a single nonmagnetic impurity in a \emph{spinful} chiral $p$-wave superconductor, where Cooper pairs have a definite angular momentum, using quasiclassical theory\cite{Volovik1999} and has found that the Majorana zero mode in the vortex core is not affected by a nonmagnetic impurity. The robustness of the Majorana fermion at zero energy has also been shown for a coreless vortex in an ``antidot'' on the surface of a disordered three-dimensional (3D) topological insulator,\cite{Ioselevich2012} where superconductivity is induced by proximity to a superconducting film with a circular hole with radius larger than the coherence length and the mean free path. Such a 3D TSC state is in the symplectic class of AII, where the topological invariant belongs to $\mathbb{Z}_2$, and the system has no particle-hole symmetry and is odd under time reversal.\cite{Schnyder2008} It is intriguing to study the effects of a nonmagnetic impurity in a 2D $s$-wave topological superconductor in the presence of vortices, which belongs to symmetry class D and has particle-hole symmetry, but no time-reversal symmetry.\cite{Schnyder2008} In this system there exist two underlying chiralities ($p_x\pm ip_y$ in the continuum model) and the effective (spinless) $p$-wave nature of the system varies depending on the material parameters. 

The purpose of the present work is to examine the effects of the two chiralities inherently present in 2D $s$-wave TSC states on vortex structure, by solving the Bogoliubov-de Gennes (BdG) equations\cite{deGennes} on the tight-binding model of Sato, Takahashi, and Fujimoto\cite{Sato2009,Sato2010} self-consistently for 
the superconducting order parameter in the vortex lattice. The tight-binding model is versatile and useful for modelling real systems in that band structure and the filling factor can easily be incorporated in terms of hopping amplitudes and chemical potential. We assume that the 2D system has Rashba SO coupling and a pairing interaction that drives $s$-wave superconductivity or superfluidity, and is under Zeeman field, e.g., generated by proximity to a ferromagnetic insulator in a heterostructure. Our model is applicable to systems such as $s$-wave superfluids of fermionic atoms created by $s$-wave Feshbach resonance in an optical lattice\cite{Sato2009}, one-atom-layer TI-Pb on Si(111)\cite{Matetskiy2015}, and ionic-liquid based electronic double-layer transistors.\cite{Li2016,Nagai2016}

Solving the BdG equations for TSC has high numerical demand as the dimension of the BdG Hamiltonian matrix is four times the total number of lattice sites, and also the system area needs to be large enough to sustain two vortices that are well separated to have a pair of isolated Majorana fermions as vortex bound states. Thus, the conventional way of solving the BdG equations by direct diagonalization is not feasible. We utilise the Chebyshev polynomial expansion method\cite{Covaci2010,Nagai2012} for solving for the order parameter self-consistently, as well as calculating the local density of states (LDOS) after self-consistency has been achieved. Furthermore, we use the numerically efficient algorithm developed by Sakurai and Sugiura \cite{Sakurai2003,Nagai2013} to obtain quasiparticle spectra within an energy window of one's choice.

The paper is organised as follows. The model is described in Sec.~\ref{sec:model}, results are presented and discussed in Sec.~\ref{sec:results}, and the work is summarised in Sec.~\ref{sec:conclusions}.

\section{\label{sec:model}Model}

We use the tight-binding model for a 2D $s$-wave topological superconductor:\cite{Sato2009,Sato2010,Nagai2015}
\begin{eqnarray} 
\mathcal{H} &=& \sum_{\langle ij \rangle \sigma} t_{ij} c^\dag_{i \sigma} c_{j \sigma} + \sum_{i\sigma} (-\mu + V_i) c^\dag_{i \sigma} c_{i \sigma}\nonumber\\
&-& h \sum_i (c^\dag_{i \uparrow} c_{i \uparrow} - c^\dag_{i \downarrow} c_{i \downarrow}) \nonumber \\
&+& \frac{\alpha}{2} \biggl[\, \sum_i (c^\dag_{i-\hat{x} \downarrow} c_{i \uparrow} - c^\dag_{i+\hat{x} \downarrow} c_{i \uparrow}) \nonumber \\
&+& i(c^\dag_{i-\hat{y} \downarrow} c_{i \uparrow} - c^\dag_{i+\hat{y} \downarrow} c_{i \uparrow}) + \text{H.c.}\, \biggr] \nonumber \\
&+& \sum_i (\Delta_{i} c^\dag_{i \uparrow} c^\dag_{i \downarrow} + \text{H.c.})\,,
\label{hamiltonian}
\end{eqnarray}
where we consider hopping among nearest-neighbour lattice sites $\langle ij \rangle$ only with the hopping amplitude $t_{ij}\equiv -t$, $\mu$ is the chemical potential, $V_i$ is the single-particle potential due to a nonmagnetic impurity at site $i$, $h$ is the Zeeman field, $\alpha > 0$ is the Rashba SO coupling strength, $\Delta_{i}$ is the $s$-wave superconducting order parameter at site $i$, and H.c. stands for the Hermitian conjugate. We set the lattice constant to be unity, and $\hat{x}$ and $\hat{y}$ are the unit vectors in the $x$ and $y$ directions. $c^\dag_{i\sigma}$ and $c_{i\sigma}$ creates and annihilates, respectively, the electron at site $i$ with spin $\sigma$ ($=\uparrow,\downarrow$). We solve the BdG equations with the Hamiltonian (\ref{hamiltonian}) and solve for the order parameter $\{\Delta_i\}$ self-consistently for a given coupling constant for the pairing interaction, $U_i\equiv U$:
\begin{equation}
\Delta_{i} = U \langle c_{i \downarrow} c_{i \uparrow} \rangle .
\end{equation}

When the system has translational symmetry, the real-space Hamiltonian in Eq.~(\ref{hamiltonian}) can be Fourier-transformed to momentum space and written as\cite{Sato2010}
\begin{equation}
\mathcal{H} = \frac{1}{2} \sum_{\bm{k}}\Psi^\dagger_{\bm{k}} \mathcal{H}(\bm{k}) \Psi_{\bm{k}}\,,
\label{hkspace}
\end{equation}
where $\Psi_{\bm{k}}=(c_{\bm{k}\uparrow}\;c_{\bm{k}\downarrow}\;c^\dagger_{-\bm{k}\uparrow}\;c^\dagger_{-\bm{k}\downarrow})^T$ and
\begin{equation}
\mathcal{H}(\bm{k}) = 
\begin{pmatrix}
	\epsilon(\bm{k}) - h\sigma_z + \alpha \mathcal{L}(\bm{k})\cdot \bm{\sigma}  & i \Delta(\bm{k}) \sigma_y \\
	-i \Delta(\bm{k})^* \sigma_y & -\epsilon(\bm{k}) + h\sigma_z + \alpha  \mathcal{L}(\bm{k})\cdot \bm{\sigma}^* 
\end{pmatrix}.
\label{hkmatrix}
\end{equation}
Here $c^\dagger_{\bm{k}\sigma}$ and $c_{\bm{k}\sigma}$ are the creation and annihilation operators of the electron with momentum $\bm{k}=(k_x,k_y)$ and spin $\sigma$, $\epsilon(\bm{k}) = -2t({\rm cos}\,k_x + {\rm cos}\,k_y) -\mu$, $\mathcal{L}(\bm{k})\equiv (\mathcal{L}_x, \mathcal{L}_y) = ({\rm sin}\,k_y, -{\rm sin}\,k_x)$, and $\bm{\sigma}\equiv (\sigma_x,\sigma_y)$ and $\sigma_z$ are the Pauli matrices.
The above Hamiltonian 
in momentum space can be diagonalized to obtain the quasiparticle spectrum as
\begin{equation}
E_\pm(\bm{k}) = \sqrt{\epsilon(\bm{k})^2 + \alpha^2 |\mathcal{L}(\bm{k})|^2 + h^2 + |\Delta|^2 \pm 2\xi(\bm{k})}\,,
\label{bulkspectrum}
\end{equation}
where 
$\xi(\bm{k}) = \sqrt{\epsilon(\bm{k})^2 \alpha^2 |\mathcal{L}(\bm{k})|^2 + (\epsilon(\bm{k})^2 + |\Delta|^2)h^2}$ and we have assumed an isotropic $s$-wave order parameter, $\Delta(\bm{k})\equiv \Delta$. Depending on the values of $\mu$, $h$, and $\Delta$, the system can be in a trivial or nontrivial (Abelian or non-Abelian) topological phase according to the first Chern number or the Thouless-Kohmoto-Nightingale-Nijs (TKNN) number,\cite{TKNN1982} which we denote as $\nu$, as classified for four different band regions in Ref.~\onlinecite{Sato2010}. The topological invariant $\nu\in\mathbb{Z}$\cite{Schnyder2008} can be calculated by\cite{Ishikawa1987,Volovik2009,Gurarie2011}
\begin{eqnarray}
\nu = &&\frac{1}{8\pi^2}\int d\bm{k} d\omega\, [\,{\rm Tr}(G\partial_{k_x}G^{-1}G\partial_{k_y}G^{-1}G\partial_{\omega}G^{-1})\nonumber\\
&&-{\rm Tr}(G\partial_{k_y}G^{-1}G\partial_{k_x}G^{-1}G\partial_{\omega}G^{-1})\,]\,,
\end{eqnarray}
where $G=(i\omega-\mathcal{H}(\bm{k}))^{-1}$.
The spectral gap $E_0$ is the minimum value of $E_\pm(\bm{k})$ in Eq.~(\ref{bulkspectrum}) and e.g., as $h$ is varied for a given set of $\alpha$, $\mu$, and $\Delta$, the system transitions from one topological (trivial or nontrivial) phase to another every time $E_0$ vanishes. The system is in Abelian and non-Abelian phase when $\nu$ is even and odd, respectively ($-2$ and $\pm 1$ in this model), and in trivial phase when $\nu=0$. Achieving non-Abelian phase with $\nu=-1$ in most of the band region $-2t < \mu \le 2t$ would require relatively large Zeeman field\cite{Sato2010,Nagai2014oval} and accordingly large values of $\alpha$ and/or $|U|$ that may be unrealistic for actual materials in order to overcome the Pauli depairing effect. Thus, in this work we focus on non-Abelian states with $\nu=\pm 1$ in the band regions $\mu \le -2t$ and $\mu > 2t$, where\cite{Sato2010}
\begin{eqnarray}
&&(4t-|\mu|)^2+\Delta^2<h^2<\mu^2+\Delta^2\,;\quad \nu = 1\,,\\
&&\mu^2+\Delta^2<h^2<(4t+|\mu|)^2+\Delta^2\,;\quad \nu = -1\,.
\end{eqnarray}

The vortex lattice is formed by uniform magnetic field applied in the $+z$ direction, $\bm{H}=H\hat{z}$. The electron wavefunction acquires the Peierls phase factor while traversing from site $j$ to $i$ (coordinate from $\bm{r}_{j}$ to $\bm{r}_{i}$) due to the associated vector potential so that the hopping amplitude is modified as
\begin{equation}
t_{ij}\,{\rm exp}\biggl[\,i\frac{e}{\hbar c} \int_{\bm{r}_{i}}^{\bm{r}_{j}} d\bm{r}\cdot\bm{A}(\bm{r}) \biggr]\,,
\end{equation}
where the vector potential $\bm{A}(\bm{r})=(\bm{H}\times \bm{r})/2$ in the symmetric gauge. To ensure obtaining Majorana fermions that come in pairs as a solution to the BdG equations,\cite{Stone2004,Sato2009p-wave} we place two vortices in our system that is the vortex unit cell, i.e., two flux quanta within the system area. The field strength is controlled by the system size as $H=2\phi_0/N_xN_y$ with $\phi_0=hc/2e$, where $N_x$ and $N_y$ are the number of lattice sites in the $x$ and $y$ directions. For the results presented below, we impose the periodic boundary condition for the vortex lattice\cite{Takigawa2000,Kawakami2016} such that there is one vortex in the centre of a square lattice and a quarter vortex at each of the four corners centred right outside the corner site, e.g., at $(x,y)=(1/2,1/2)$. For all the results shown, an odd number of lattice sites $N_x=N_y$ has been used so that the vortex inside the system is centred at the centre site of the square lattice. 
We have used the convention that the electron charge is negative ($e>0$) and hence the quasiparticle bound states carry the angular momentum of $-\hbar$ about the vortex centre. The phase winding of the order parameter is referred to as the vorticity ($-1$ in our calculation) hereafter.

To study the effects of nonmagnetic impurities, we place a single nonmagnetic impurity with potential $V_i\equiv V_{\rm imp}$ at the centre site of the system. This is where one of the vortices is centred at, which is a reasonable assumption as a vortex tends to be pinned by an impurity or a defect in real materials, and $V_{\rm imp}$ can be thought of as a pinning potential. We stop self-consistent iterations for the order parameter at the $l$-th iteration step, when the order parameter as a complex vector $\vec{\Delta}$ of length $N_xN_y$ satisfies
\begin{equation}
\frac{\|\vec{\Delta}^{(l)}-\vec{\Delta}^{(l-1)}\|}{\|\vec{\Delta}^{(l-1)}\|} < \delta\,,
\end{equation}
where we set $\delta=10^{-6}$. We have found that results do not change with tighter convergence criteria with $\delta$ as small as $10^{-10}$.

\begin{figure}
  \begin{center}
    \begin{tabular}{p{0.95\columnwidth}}
      (a)\resizebox{!}{!}{\includegraphics[width=0.9\columnwidth]{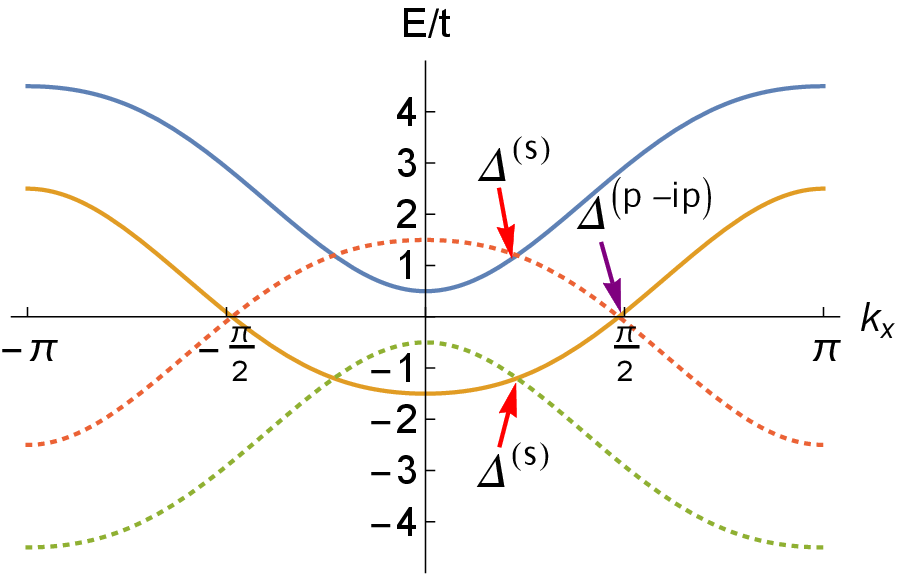}}\\
      \\
      (b)\resizebox{!}{!}{\includegraphics[width=0.9\columnwidth]{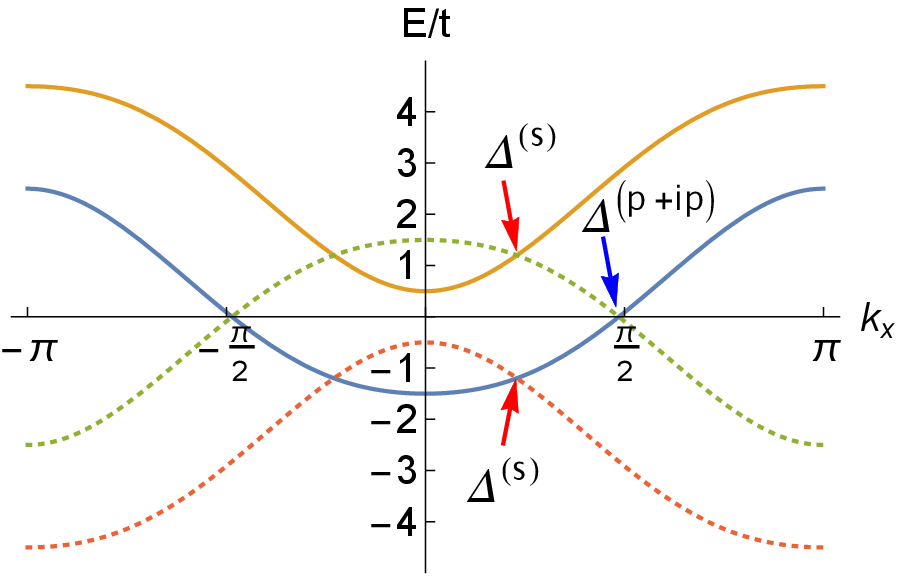}}
    \end{tabular}
  \end{center}
  \caption{\label{fig:bands} (Colour online) Electron (solid curves) and hole (dashed curves) bands $\pm E_{\pm}$ in the normal state for $\mu=-3.5t$, $\alpha=t$, and $k_y=0$ as a function of $k_x$ for (a) $h=t$ and (b) $h=-t$, where the Fermi surface is formed by $E_-$ (orange curve) and $E_+$ (blue curve), respectively. The interband (intraband) pairing is represented by $\Delta^{({\rm s})}$ ($\Delta^{({\rm p}\pm i{\rm p})}$ corresponding to chiralities $\eta_\pm$). 
  }
\end{figure}

The energy spectrum (\ref{bulkspectrum}) depends only on the magnitude of the Zeeman field $h$ and so the phase transition between different topological phases is independent of the sign of $h$. In a given topological phase, however, which one of the two underlying chiralities, $\eta_{\pm}\equiv -({\cal L}_x\pm i{\cal L}_y)/\sqrt{{\cal L}_x^2+{\cal L}_y^2}=\pm i({\rm sin}\,k_x \pm i\,{\rm sin}\,k_y)/\sqrt{{\rm sin}^2k_x+{\rm sin}^2k_y}$,\cite{Fujimoto2008,Sato2009p-wave} is more manifest is determined by the sign of $h$\cite{Wu2012} as well as the sign of $\mu$.\cite{Sato2010} This can be seen by expressing $\mathcal{H}(\bm{k})$ in the ``chirality basis'' that diagonalizes the normal-state Hamiltonian $\epsilon(\bm{k}) - h\sigma_z + \alpha \mathcal{L}(\bm{k})\cdot \bm{\sigma}$:\cite{Sato2010}
\begin{equation}
\tilde{\mathcal{H}}(\bm{k}) = 
\begin{pmatrix}
	\epsilon(\bm{k}) + \Delta\epsilon(\bm{k})\sigma_z & \hat{\Delta} \\
	\hat{\Delta}^\dagger & -\epsilon(\bm{k}) - \Delta\epsilon(\bm{k})\sigma_z
\end{pmatrix},
\label{hktilde}
\end{equation}
where $\Delta\epsilon(\bm{k}) = {\rm sgn}(h) \sqrt{ \alpha^2 |\mathcal{L}(\bm{k})|^2 + h^2}$ and
\begin{equation}
\hat{\Delta} = \frac{1}{\Delta\epsilon(\bm{k})}
\begin{pmatrix}
	-\alpha|\mathcal{L}(\bm{k})|\eta_+\Delta(\bm{k}) & h\Delta(\bm{k}) \\
	-h\Delta(\bm{k}) & -\alpha|\mathcal{L}(\bm{k})|\eta_-\Delta(\bm{k}) 
\end{pmatrix}.
\label{deltahat}
\end{equation}
In the normal state, the eigenspectrum in Eq.~(\ref{bulkspectrum}) reduces to
$E_\pm(\bm{k}) = \epsilon(\bm{k})\pm \Delta\epsilon(\bm{k})$, 
where ${\rm sgn}(h)$ in $\Delta\epsilon(\bm{k})$ is necessary for obtaining the correct eigenvalues and eigenfunctions as can be seen by taking the limit $\alpha\rightarrow 0$. We note that this factor is missing in the discussion of the chirality basis in Ref.~\onlinecite{Sato2010} where $h$ was supposed to be positive. It can be seen in Eq.~(\ref{deltahat}) that the \emph{intraband} pairing in the band $E_+$ ($E_-$) has the chirality of $\eta_+$ ($\eta_-$), while the interband pairing is purely $s$-wave. This is analogous to the two chiralities $p_x\pm ip_y$ present in the non-Abelian phase of the continuum model.\cite{Alicea2010,Shitade2015}. Although orbital angular momentum is not a good quantum number in the presence of SO coupling, it has been found in Ref.~\onlinecite{Shitade2015} that when $\alpha$ is small enough, the average angular momentum carried by a Cooper pair can be close to $-\hbar$ in non-Abelian states dominated by $p_x-ip_y$, as in chiral $p$-wave superconductors. 

\begin{figure}
  \includegraphics[width=0.9\columnwidth]{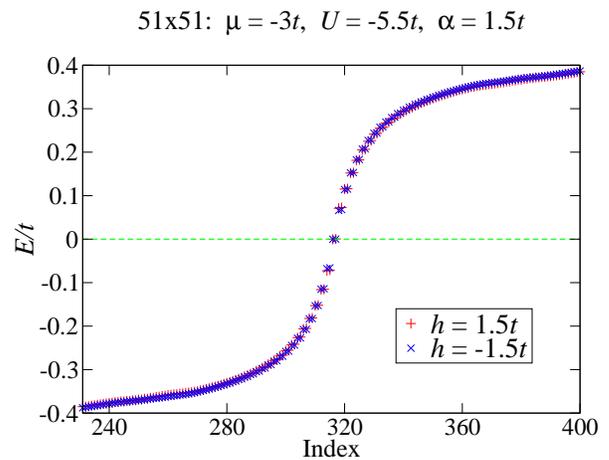}
  \caption{\label{fig:eigmu-3} (Colour online) Quasiparticle excitation spectra for $h=1.5t$ (red plus) and $h=-1.5t$ (blue cross) in the vortex lattice in a pure 51$\times$51 system for $\mu=-3t$, $U=-5.5t$, $\alpha=1.5t$. The index numbers the eigenvalues.
  }
\end{figure}

We illustrate in Fig.~\ref{fig:bands} for $\mu=-3.5t$ and $\alpha=t$ the two cases, (a) $h=t$ and (b) $h=-t$, where the normal-state Fermi surface is formed by $E_-$ and $E_+$, respectively. This value of $\alpha$ is small enough so that the respective chirality, $\eta_-$ and $\eta_+$, associated with the Fermi surface can be dominant over the other, as discussed in Sec.~\ref{sec:chirality} for $\mu=3.5t$ (where the dominant chirality is $\eta_+$ and $\eta_-$, respectively, for $h=t$ and $h=-t$). Due to the interband pairing, however, both chiralities are present in a TSC state in general. We will demonstrate in Secs.~\ref{sec:chirality} and \ref{sec:weak-alpha} that for relatively weak SO coupling, vortex structure and effects of a nonmagnetic impurity can be influenced strongly by one of the chiralities $\eta_+$ and $\eta_-$, which we call the chirality of $+1$ and $-1$, respectively, in the remainder of the paper.

The major chirality in a non-Abelian TSC state stems from the chirality (spin winding) of the single Fermi surface in the normal state.\cite{Sato2010,Alicea2010,Shitade2015}. The latter is $+1$ for $\mu>0$, $h>0$ and $\mu<0$, $h<0$, and $-1$ for $\mu>0$, $h<0$ and $\mu<0$, $h>0$: The first and second $\mu<0$ combinations correspond to the convention used in the continuum model in Refs.~\onlinecite{Masaki2014}, \onlinecite{Masaki2015} and Ref.~\onlinecite{Shitade2015}, respectively. In the case of a trivial TSC state, the chirality of the dominant one of the two Fermi surfaces can manifest itself. The TKNN number $\nu$ reverses its sign under $h\rightarrow -h$, whereas it does not under $\mu\rightarrow -\mu$. Whether trivial or nontrivial, the underlying chirality in the TSC state -- rather than the TKNN number itself -- governs how the system responds to vorticity and nonmagnetic impurities in the presence of a vortex.

\section{\label{sec:results}Results}

\subsection{\label{sec:chirality}Chirality vs Vorticity}

\begin{figure}
\includegraphics[width=0.95\columnwidth]{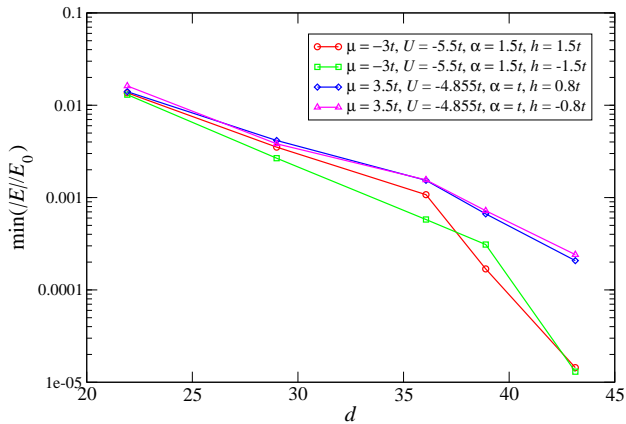}
\caption{\label{fig:zero} (Colour online) Minimum quasiparticle energy as a function of the distance between nearest-neighbour vortices $d=N_x/\sqrt{2}$ for $N_x=N_y=31$, 41, 51, 55, and 61.
}
\end{figure}

\begin{figure}
  \includegraphics[width=0.95\columnwidth]{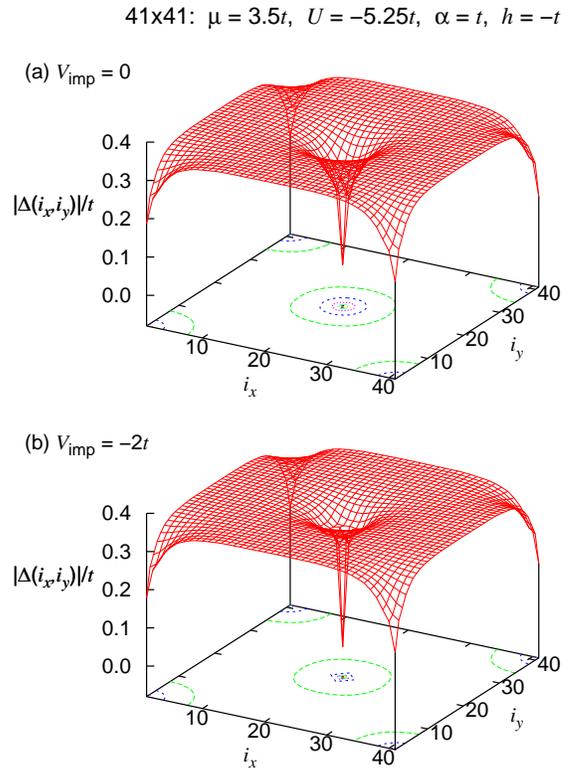}
  \caption{\label{fig:orderU-5.25} (Colour online) Order parameter of the vortex lattice in a 41$\times$41-site system for $\mu=3.5t$, $U=-5.25t$, $\alpha=t$, and $h=-t$, with (a) $V_{\rm imp}=0$ and (b) $V_{\rm imp}=-2t$ at the lattice centre.
  }
\end{figure}

\begin{figure}
  \includegraphics[width=0.9\columnwidth]{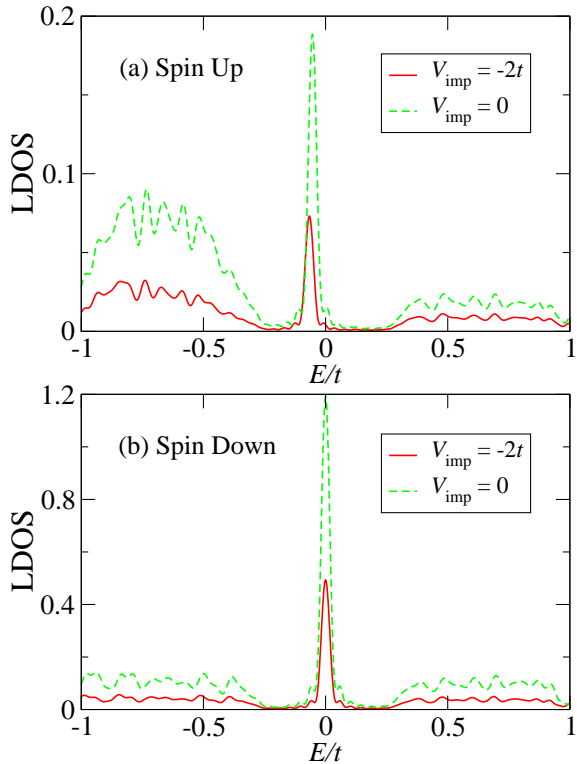}
  \vspace{.2cm}
  \caption{\label{fig:ldosU-5.25_h1} (Colour online) (a) Spin-up and (b) spin-down components of the LDOS as a function of quasiparticle energy at the centre of the 41$\times$41 vortex lattice for $\mu=3.5t$, $U=-5.25t$, $\alpha=t$, and $h=t$; for $V_{\rm imp}=-2t$ (red solid curves) and $V_{\rm imp}=0$ (green dashed curves).
  }
\end{figure}

\begin{figure}
  \includegraphics[width=0.9\columnwidth]{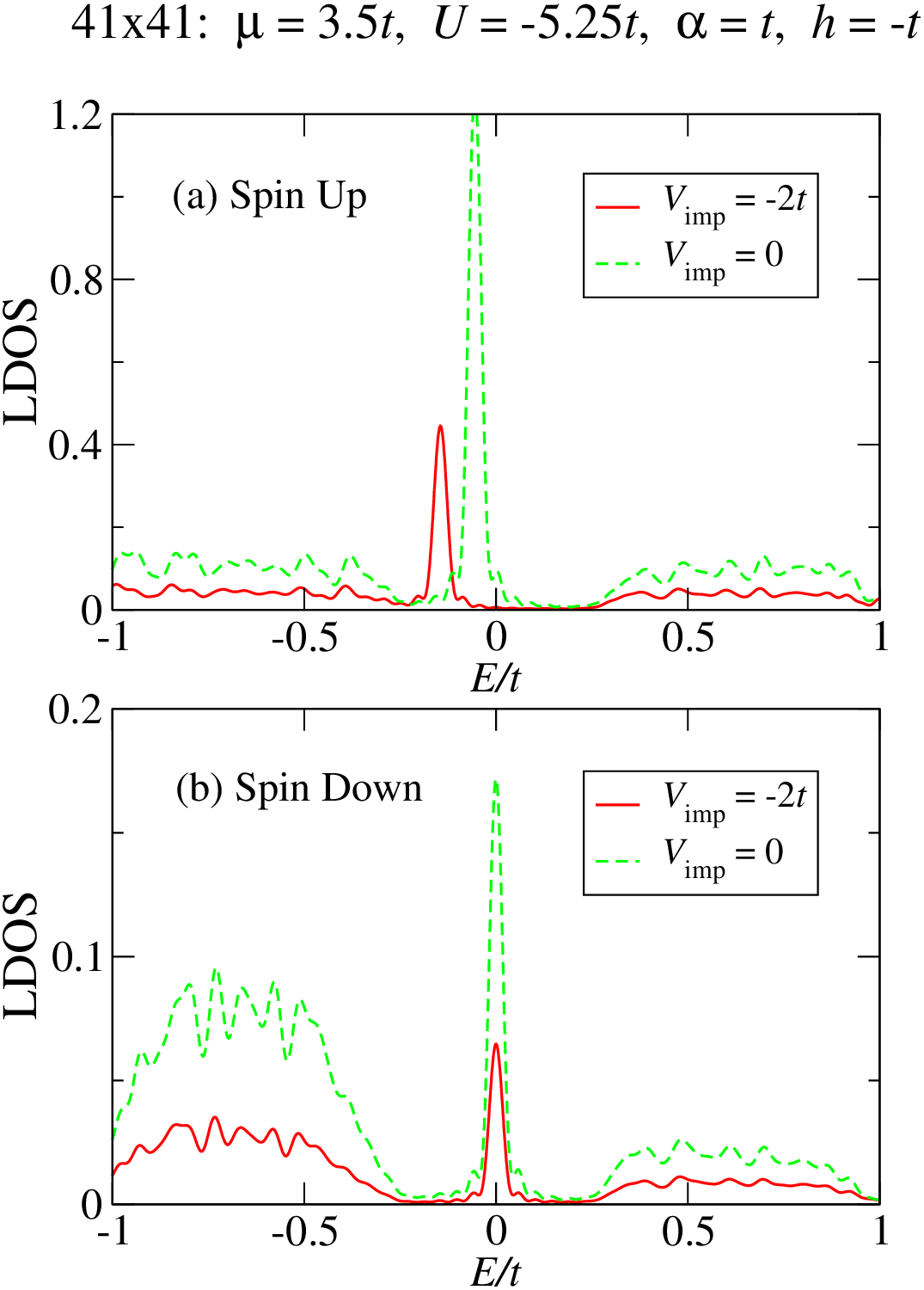}
  \vspace{.2cm}
  \caption{\label{fig:ldosU-5.25_h-1} (Colour online) Same as Fig.~\ref{fig:ldosU-5.25_h1}, but for $h=-t$.
  }
\end{figure}

We have studied the vortex lattice states for a wide variety of parameter sets ($\mu$, $\alpha$, $h$, $U$ and the resulting bulk order parameter $\Delta$ and spectral gap $E_0$) within our tight-binding model (\ref{hamiltonian}), in particular exploring the effects of the strength of SO coupling $\alpha$ and the sign of the Zeeman field $h$. We find that for $\alpha\gtrsim t$ the sign of $h$ hardly makes any difference in a pure vortex lattice state (i.e., with no impurity). The converged order parameter, the excitation spectrum, and the ground-state energy are practically the same for positive and negative $h$ in the absence of impurity. We will discuss the influence of chirality on the vortex structure that can be apparent even in a pure system for relatively small $\alpha$ in Sec.~\ref{sec:weak-alpha}.

\begin{figure}
  \includegraphics[width=0.9\columnwidth]{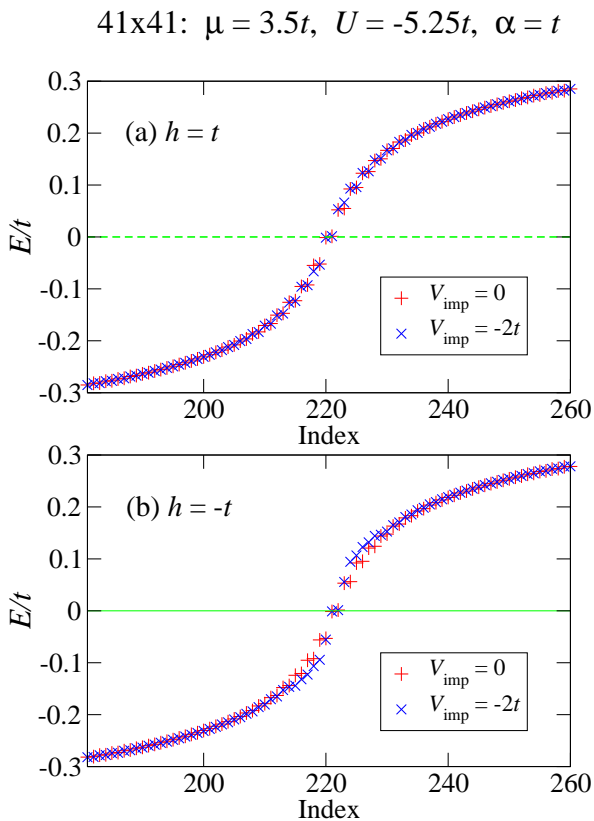}
  \vspace{.2cm}
  \caption{\label{fig:eigU-5.25} (Colour online) Excitation spectra in the 41$\times$41 vortex lattice for $\mu=3.5t$, $U=-5.25t$, $\alpha=t$, comparing the cases with $V_{\rm imp}=0$ (red plus) and $V_{\rm imp}=-2t$ (blue cross) at a vortex centre for (a) $h=t$ and (b) $h=-t$. The index numbers the eigenvalues.
  }
\end{figure}

In Fig.~\ref{fig:eigmu-3} we show the quasiparticle spectra for $h=\pm 1.5t$ in the vortex lattice in a pure 51$\times$51-site system for $\mu=-3t$, $\alpha=1.5t$, and $U=-5.5t$ ($\Delta\simeq 0.5t$, $E_0\simeq 0.33t$, and $\nu=1$), where the abscissa is an index numbering the discrete eigenvalues. The major change caused by reversal of the Zeeman field while keeping all other parameters the same is the average numbers of spin-up and spin-down electrons in the system being interchanged. Accordingly, the overall magnitude of the spin-up and spin-down probability amplitudes of the Majorana bound state (and low-energy quasiparticle excitations) in each vortex core changes when the sign of $h$ is flipped: e.g., if the spin-up component of a Majorana is dominant over the spin-down component for a given $h$, then vice versa for $-h$. Other than this, the overall structure of each of the spin-up and spin-down Majorana wave functions is little affected by the direction of the Zeeman field. It is just barely discernible in Fig.~\ref{fig:eigmu-3} that the first excited state has a slightly lower energy for $h=-1.5t$ than for $h=1.5t$. We find that the opposite is true for positive $\mu$, though the difference tends to be very small. The Majorana bound-state energy is $\sim 0.0003t$ and $\sim 0.0002t$, respectively, for $h=1.5t$ and $h=-1.5t$ in this example. These energy levels are not exactly zero due to nonzero overlap of the Majorana wavefunctions bound to nearest-neighbour vortices on the finite lattice. To demonstrate that they are indeed Majorana bound states, the minimum eigenvalue $|E|$ in units of the bulk spectral gap $E_0$ is plotted in Fig.~\ref{fig:zero} as a function of the distance between nearest-neighbour vortices for $N_x=N_y=31$, 41, 51, 55, and 61; for the two systems shown in Fig.~\ref{fig:eigmu-3} and for $\mu=3.5t$, $\alpha=t$, $h=\pm 0.8t$, and $U=-4.855t$ ($\Delta\simeq 0.37t$, $E_0\simeq 0.18t$, and $\nu=1$). It can be seen that the minimum eigenvalue approaches zero as the system size increases.

\begin{figure}
  \includegraphics[width=0.7\columnwidth]{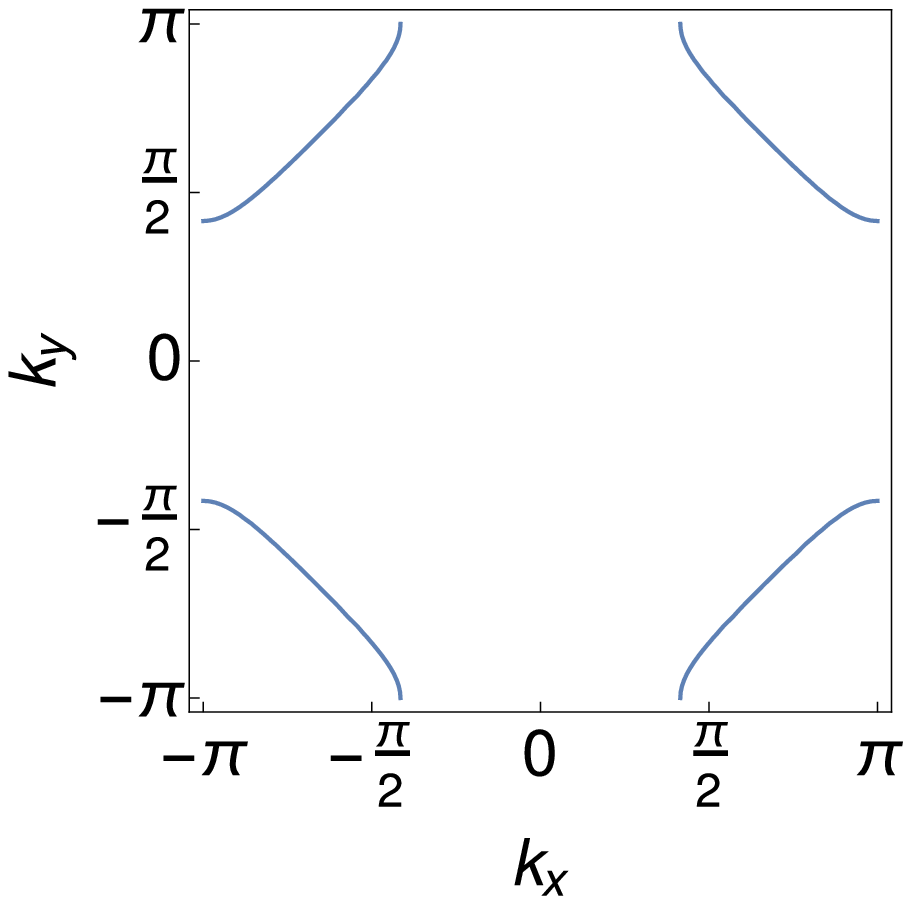}
  \caption{\label{fig:FS_alpha2.5} (Colour online) Normal-state Fermi surface $E_-$ for $\mu=-2t$, $\alpha=2.5t$ and $h=2.5t$ as a function of $k_x$ and $k_y$.
  }
\end{figure}

Presented in Fig.~\ref{fig:orderU-5.25} is the magnitude of the converged order parameter $\Delta(i_x,i_y)$ of the vortex lattice in a 41$\times$41-site system for $\mu=3.5t$, $U=-5.25t$, $\alpha=t$, $h=-t$ ($\Delta\simeq 0.37t$ and $\nu=1$) as a function of $x\equiv i_x$ and $y\equiv i_y$ coordinates; with (a) no impurity and (b) a single nonmagnetic impurity with potential $V_{\rm imp}=-2t$ at the centre of the lattice, where one of the vortices is centred at. The contour projection of the order parameter magnitude onto the $xy$ plane is in steps of $0.05t$. For this value of $\alpha$, the vortex structure in a pure system barely depends on the sign of $h$, and $|\Delta(i_x,i_y)|$ for $h=+t$ is very similar to the one shown in Fig.~\ref{fig:orderU-5.25}(a) for $h=-t$. In the presence of a nonmagnetic impurity, however, one of the two chiralities can manifest itself in the vortex core structure. For $V_{\rm imp}=-2t$, while the vortex structure of the order parameter does not change much from the pure case for $h=+t$ (not shown), it can be seen clearly in Fig.~\ref{fig:orderU-5.25}(b) that for $h=-t$, the vortex core shrinks in comparison to the no-impurity case shown in Fig.~\ref{fig:orderU-5.25}(a). 

\begin{figure}
  \includegraphics[width=0.95\columnwidth]{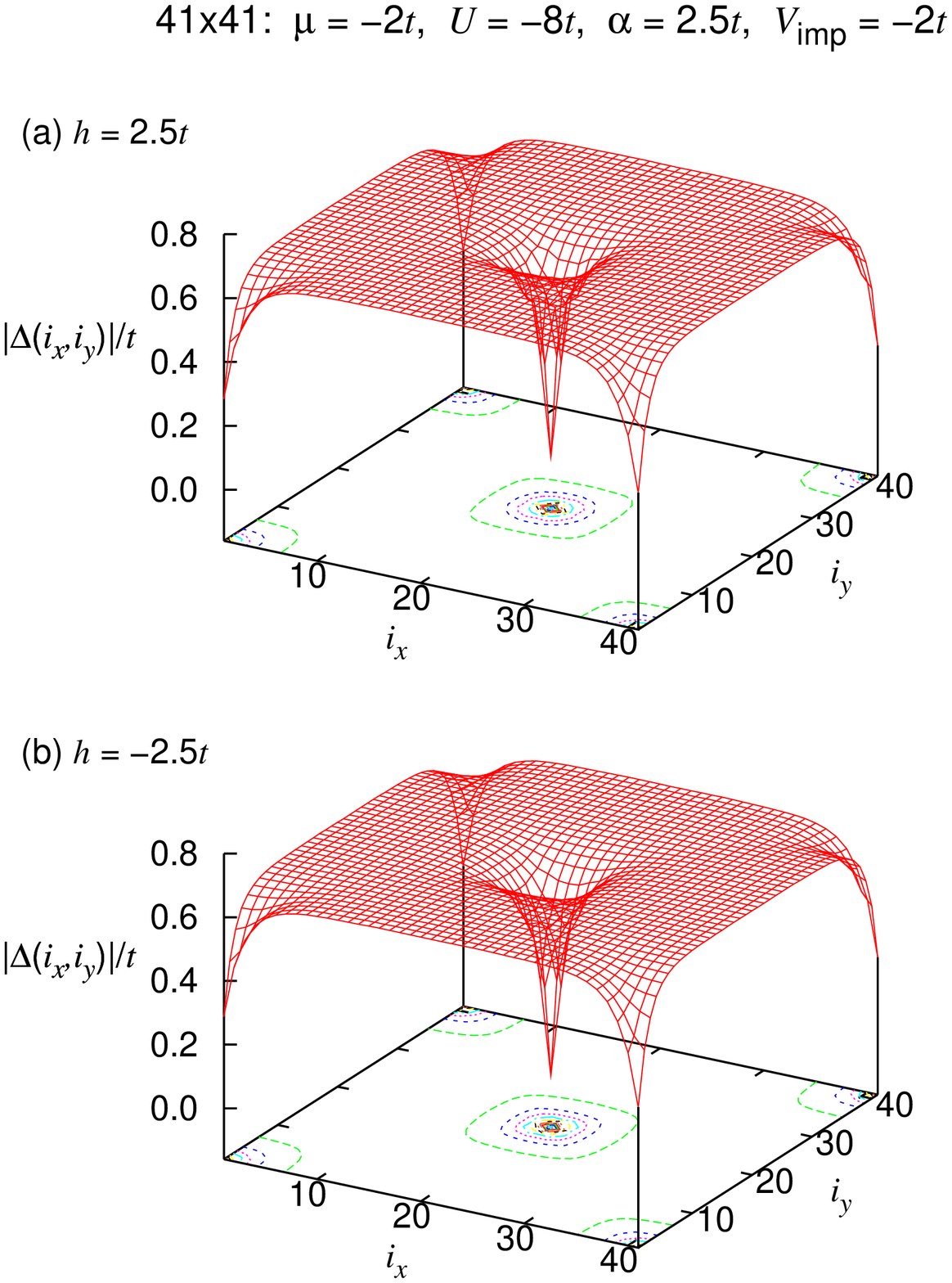}
  \caption{\label{fig:orderU-8} (Colour online) Order parameter of the vortex lattice in a 41$\times$41-site system for $\mu=-2t$, $U=-8t$, $\alpha=2.5t$, and $V_{\rm imp}=-2t$ at the lattice centre, for (a) $h=2.5t$ and (b) $h=-2.5t$. Contour projection in steps of $0.05t$.
  }
\end{figure}

The LDOS at the vortex centre (where the nonmagnetic impurity is placed at) is plotted as a function of quasiparticle energy in Fig.~\ref{fig:ldosU-5.25_h-1} for the system presented in Fig.~\ref{fig:orderU-5.25} and in Fig.~\ref{fig:ldosU-5.25_h1} for the same system but for $h=+t$, where the (a) spin-up and (b) spin-down components of the LDOS are compared between $V_{\rm imp}=0$ and $-2t$. The Lorentz kernel\cite{Weisse2006,Covaci2010} has been used for calculation of the LDOS with the corresponding Lorentzian smoothing width of $0.0005t$. The number of terms summed over in the Chebyshev polynomial expansion of the Green function components for the LDOS is 3000, while the cutoff of 2000 terms was used for the order parameter. The bulk spectral gap in this system (irrespective of the sign of $h$) is $E_0\simeq 0.26t$. We first note that the Majorana bound state is dominated by its spin-down component at the vortex centre for both $h=\pm t$ [Figs.~\ref{fig:ldosU-5.25_h1}(b) and \ref{fig:ldosU-5.25_h-1}(b)]. It turns out to be always the case for $\alpha\gtrsim t$ that spin of the Majorana zero mode is dominantly down at the vortex centre, regardless of the sign of the Zeeman field or the chemical potential. As mentioned above in regard to the Zeeman field, one spin component can be dominant over the other in a Majorana for a given $h$ and $\mu$, and the dominance is reversed under the change $h\rightarrow -h$, or $\mu\rightarrow -\mu$. Whether up spin or down spin is dominant in the wave function overall, spin of the Majorana bound state is mostly down at the vortex centre for $\alpha\gtrsim t$.

This spin polarization at the vortex centre is consistent with the findings of Ref.~\onlinecite{Nagai2014top}, where Majorana bound states in a three-dimensional topological superconductor with spin-orbit coupling have been found to be spin-polarised in a vortex core. Assuming circular symmetry around the vortex centre, the spin-up and spin-down wavefunctions are given by the Bessel function of the first kind, $J_n(r)$ and $J_{n-1}(r)$, respectively, where $n$ and $n-1$ are the orbital angular momentum carried by the spin-up and spin-down components of a quasiparticle, for the chirality of $+1$.\cite{Nagai2014top} The Majorana bound state has either $n=0$ or $n=1$, whichever yields the lowest (zero) energy, depending on the vorticity. In Ref.~\onlinecite{Nagai2014top} the vorticity of $+1$ was used that resulted in the Majorana wavefunction dominated by its spin-up component in the vortex core.

It can be seen in Fig.~\ref{fig:ldosU-5.25_h1} that both spin-up and spin-down (Majorana) bound-state energies are hardly affected by the impurity for $h=+t$. This can be understood in terms of the major chirality being $+1$: the angular momentum $+\hbar$ carried by Cooper pairs in the condensate is mostly cancelled by the angular momentum $-\hbar$ carried by the supercurrent, making the system effectively $s$-wave-like in the vortex core region. In contrast, the spin-up (non-Majorana) bound-state energies are shifted substantially by the presence of the impurity for $h=-t$ [Fig.~\ref{fig:ldosU-5.25_h-1}(a)], where the chirality has the same sign as the vorticity. Also in this case, however, the Majorana bound state is robust in that its energy is unchanged, albeit with reduced amplitude at the impurity site [Fig.~\ref{fig:ldosU-5.25_h-1}(b)].

Figures~\ref{fig:eigU-5.25}(a) and \ref{fig:eigU-5.25}(b) show the quasiparticle spectra for the systems presented in Figs.~\ref{fig:ldosU-5.25_h1} and \ref{fig:ldosU-5.25_h-1}, respectively. It can be seen that even for $h=t$, the first excited state is pushed up in energy though slightly, resulting in an increased minigap, by the nonmagnetic impurity at the vortex centre. Note that each energy level is doubly degenerate as there are two vortices in the lattice, though numerically not exactly as the two vortices are not equivalent: Only one of them has its centre in the system, where we place a single nonmagnetic impurity, hence shifting the energy of only one of the two states (in each level). The only perceptible change caused by the impurity for $h=t$ is the small shift of the first excited level, and this is reflected in the LDOS in Fig.~\ref{fig:ldosU-5.25_h1}(a) as the slight shift in the peak position. For $h=-t$, on the other hand, the first few excited states are affected by the impurity, and this results in the substantial shift of the bound-state peak in the LDOS at the vortex centre, as can be seen in Fig.~\ref{fig:ldosU-5.25_h-1}(a). 

For both signs of $h$, the energy of the Majorana bound state remains practically the same ($\sim 0.001t$ on the 41$\times$41 lattice) with or without the impurity. This turns out to be the case also for stronger impurity potential, or stronger spin-orbit coupling as discussed in Sec.~\ref{sec:strong-alpha}, where the vortex bound states are significantly affected by the impurity for $h>0$ as well as for $h<0$. We also find that the minigap is increased by the presence of a nonmagnetic impurity at the vortex centre, regardless of the (nonzero) magnitude, or the sign of the potential -- in case of a positive potential, as long as the magnitude is large enough so that the order parameter is suppressed at the impurity site in the absence of vortices. The order parameter can be enhanced at the impurity site for a relatively weak, positive potential,\cite{Hu2013,Goertzen2016} and the BdG equations tend to have difficulty converging when such an impurity is placed at the vortex centre, especially for $h>0$ presumably due to competition between the dominant chirality and the vorticity. By including a Gaussian impurity potential in the core of a single vortex in the continuum model,\cite{Sau2010} the authors of Ref.~\onlinecite{Mao2010} showed that the minigap simply increased continuously as the magnitude of the positive Gaussian potential was increased from zero. This result may have been possible because these authors did not solve the BdG equations self-consistently for the vortex state within the topological superconductor.

\begin{figure}
  \includegraphics[width=0.9\columnwidth]{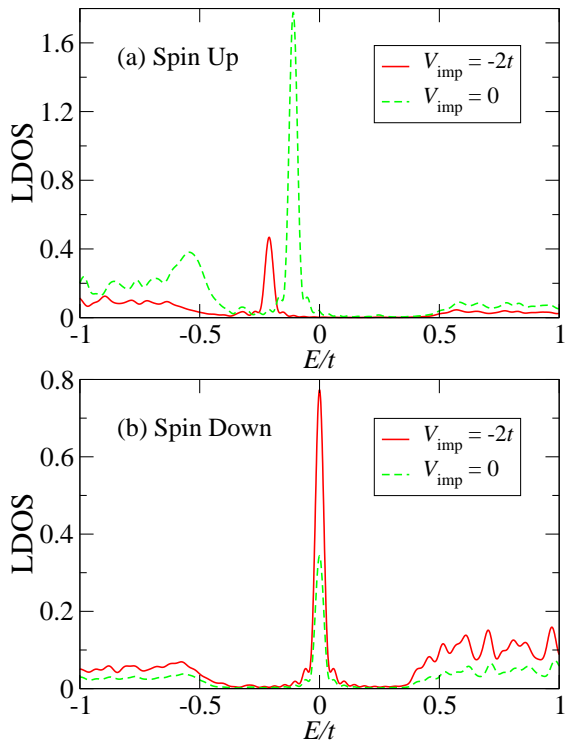}
  \vspace{.2cm}
  \caption{\label{fig:ldosU-8_h2.5} (Colour online) (a) Spin-up and (b) spin-down components of the LDOS as a function of quasiparticle energy at the vortex centre for the system presented in Fig.~\ref{fig:orderU-8}(a) for $h=2.5t$ (red solid curves) in comparison with the pure case (green dashed curves).
  }
\end{figure}

\begin{figure}
  \includegraphics[width=0.9\columnwidth]{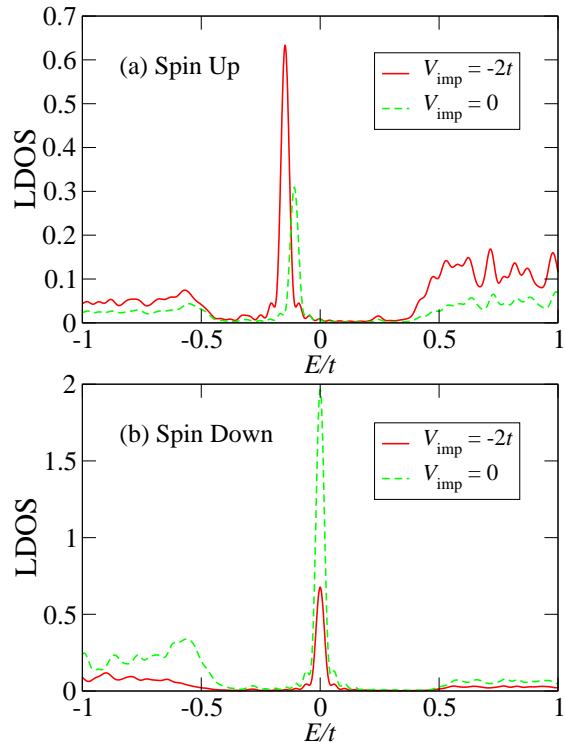}
  \vspace{.2cm}
  \caption{\label{fig:ldosU-8_h-2.5} (Colour online) Same as Fig.~\ref{fig:ldosU-8_h2.5}, but for $h=-2.5t$.
  }
\end{figure}

\begin{figure}
  \includegraphics[width=0.9\columnwidth]{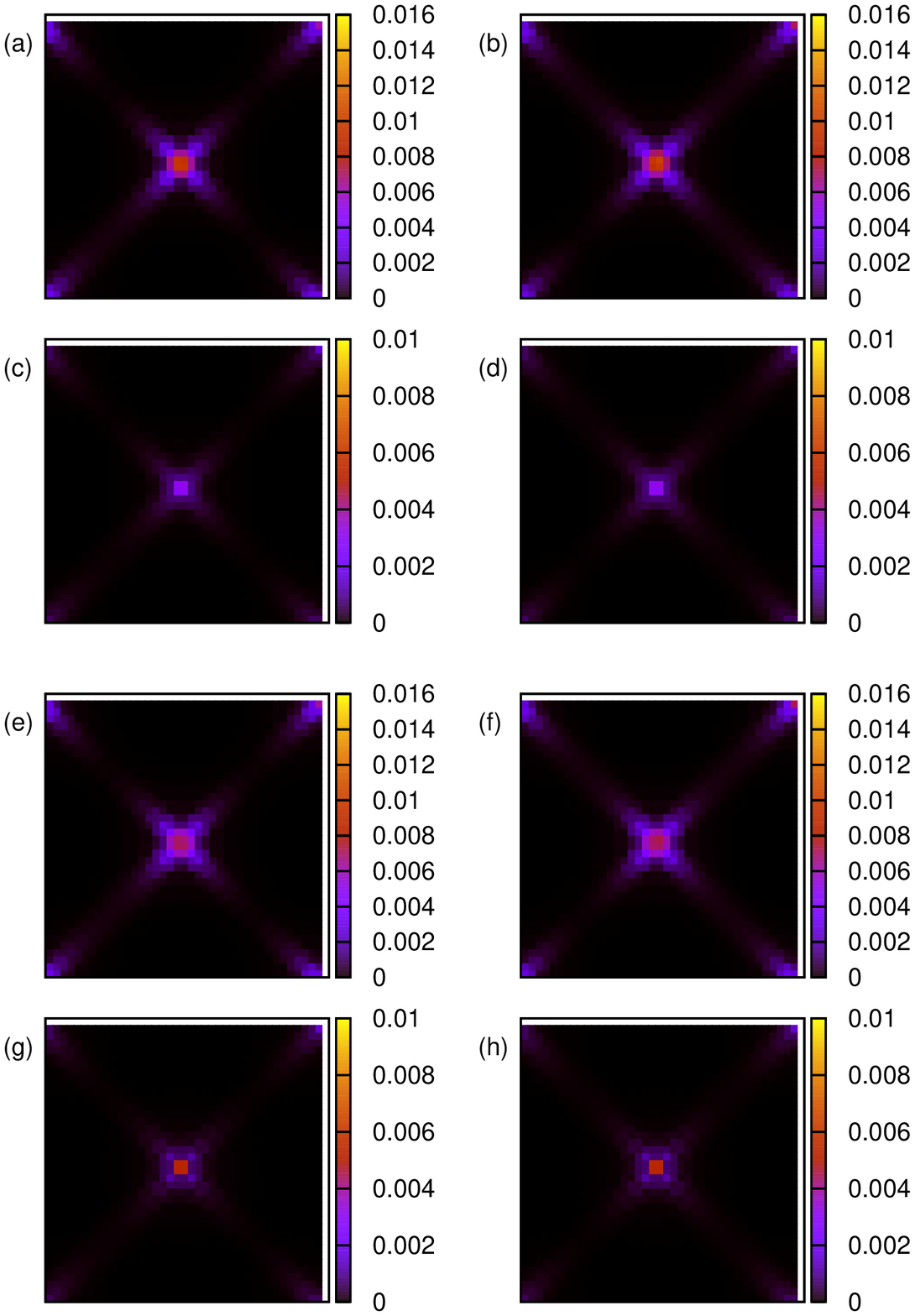}
  \vspace{.2cm}
  \caption{\label{fig:wfU-8_h2.5} (Colour online) Particle (left column) and hole (right column) probability amplitudes of the Majorana bound state for the system shown in Fig.~\ref{fig:ldosU-8_h2.5} for $h=2.5t$; for spin up (a),(b) and spin down (c),(d) for $V_{\rm imp}=0$, and for spin up (e),(f) and spin down (g),(h) for $V_{\rm imp}=-2t$.
  }
\end{figure}

\begin{figure}
  \includegraphics[width=0.9\columnwidth]{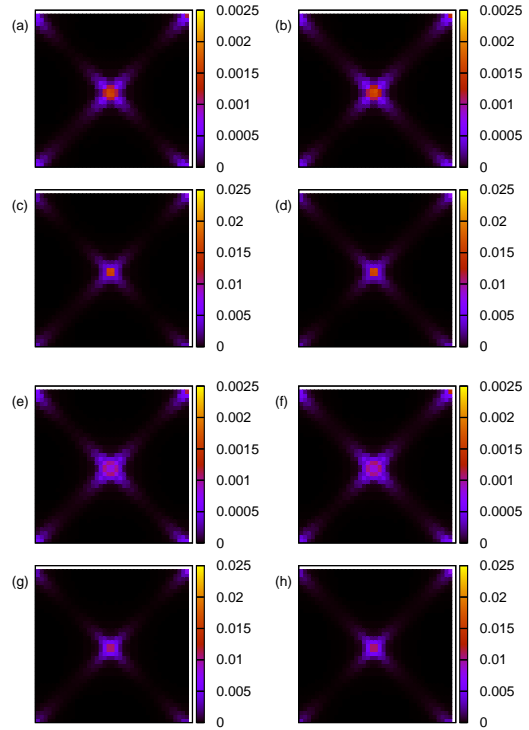}
  \vspace{.2cm}
  \caption{\label{fig:wfU-8_h-2.5} (Colour online) Same as Fig.~\ref{fig:wfU-8_h2.5}, but for $h=-2.5t$.
  }
\end{figure}

\subsection{\label{sec:strong-alpha}Strong SO Coupling}

For the systems studied in the previous subsection, the chemical potential $\mu=3.5t$ is fairly close to the top of the band and also the SO coupling constant $\alpha=t$ is relatively small so that the Fermi surface in the normal state is nearly perfectly circular. This means that the system has rotational symmetry to a good approximation and the difference between the two intrinsic chiralities can be apparent in vortex states. In this subsection, we examine a case where the chemical potential is far away enough from the top or bottom of the band and/or SO coupling is relatively strong so that the Fermi surface is nowhere close to having circular symmetry. For this purpose we use $\mu=-2t$ and $\alpha=h=2.5t$. The normal-state Fermi surface $E_-$ is shown in Fig.~\ref{fig:FS_alpha2.5}, which is centred at $(\pi,\pi)$ like the Fermi surface ($E_+$) for $\mu=3.5t$ and $\alpha=h=t$ (not shown). One can see that a large portion of this Fermi surface is almost flat and perpendicular to the directions $k_y=\pm k_x$. 

The converged order parameter for $\mu=-2t$, $\alpha=2.5t$, and $U=-8t$ ($\Delta\simeq 0.67t$, $E_0\simeq 0.39t$, and $\nu=-1$) in the 41$\times$41 vortex lattice with $V_{\rm imp}=-2t$ at the centre of the system is shown in Fig.~\ref{fig:orderU-8} for (a) $h=2.5t$ and (b) $h=-2.5t$. The contour projection onto the $xy$ plane is again in steps of $0.05t$. It can be seen that the vortex core size is not much different for the positive and negative $h$ in this case: It also turns out that it does not change much from the pure case ($V_{\rm imp}=0$) for either sign of $h$. Furthermore, in contrast to the results for $\mu=3.5t$, $\alpha=t$, and $h=-t$ in Fig.~\ref{fig:orderU-5.25}, the contour projection shows that the order parameter has more square-like symmetry rather than circular about the vortex centre. We note that it is more so for negative $h$ than positive $h$, and even in the $\mu=3.5t$ result for $h=-t$ shown in Fig.~\ref{fig:orderU-5.25} the order parameter has square symmetry ($45^\circ$ rotated with respect to the $xy$ axes) very close to the vortex centre. 

Figures~\ref{fig:ldosU-8_h2.5} and \ref{fig:ldosU-8_h-2.5} present the (a) spin-up and (b) spin-down components of the LDOS at the vortex centre as a function of quasiparticle energy for the systems shown in Fig.~\ref{fig:orderU-8}, in comparison with the pure case. The influence of the impurity is visible for both $h=\pm 2.5t$, with some energy shifts for spin-up quasiparticles. In fact, vortex bound states are more affected by the impurity for positive $h$ than for negative $h$: the shift of the spin-up LDOS peak for $h=2.5t$ is a reflection of the first excited state moved up in energy from $\sim 0.11t$ to $\sim 0.21t$, thus increasing the minigap by about $0.1t$. Interestingly, unlike the results for $\mu=3.5t$, $\alpha=t$, and $h=\pm t$ in Figs.~\ref{fig:ldosU-5.25_h1} and \ref{fig:ldosU-5.25_h-1}, the Majorana bound state for $h=2.5t$ and some of the spin-up bound states for $h=-2.5t$ become more bound to the vortex centre by the presence of the impurity. Yet, the energy of the Majorana bound state ($\lesssim 0.003t$ in this example) is barely changed by the impurity for both signs of $h$.

We show in Figs.~\ref{fig:wfU-8_h2.5} and \ref{fig:wfU-8_h-2.5} the Majorana wave functions for the systems presented in Figs.~\ref{fig:ldosU-8_h2.5} and \ref{fig:ldosU-8_h-2.5}, respectively. In each of Figs.~\ref{fig:wfU-8_h2.5} and \ref{fig:wfU-8_h-2.5}, the spin-up (a) particle and (b) hole probability amplitudes and the spin-down (c) particle and (d) hole probability amplitudes for $V_{\rm imp}=0$ are plotted as a function of $x$ and $y$ coordinates for the entire 41$\times$41-site system. Respective probability amplitudes are plotted for $V_{\rm imp}=-2t$ in (e) and (f), and (g) and (h). We first note that the particle and hole probability amplitudes in a given spin component are practically identical for all the cases shown, as expected for a Majorana fermion. Fourfold-symmetric extension of the wave functions can be discerned especially for the spin-up components (and the spin-down components for $h=-2.5t$ and $V_{\rm imp}=-2t$), reflecting $k_y=\pm k_x$ in the Fermi wave vector. The bound-state energy for this finite-size system being slightly higher than the counterpart for $\mu=3.5t$, $\alpha=t$, and $h=\pm t$ mentioned at the end of Sec.~\ref{sec:chirality} can be attributed to the nonzero overlap of the Majorana wave functions among the nearest-neighbour vortices.

Masaki and Kato have found\cite{Masaki2014,Masaki2015} in terms of the continuum model\cite{Sau2010} that for weak SO coupling, vortex bound states are more robust against nonmagnetic impurities when the chirality is opposite to the vorticity, compared to the case where the chirality and vorticity are in the same direction, and that this difference between the two chiralities diminishes as SO coupling is made stronger. They have also found that the scattering rates of zero-energy bound states are very small regardless of the major chirality and the strength of SO coupling.\cite{Masaki2015} Our results, though only with a single nonmagnetic impurity at the vortex centre, are consistent with their findings.

\subsection{\label{sec:weak-alpha}Weak SO Coupling}

\begin{figure}
  \includegraphics[width=0.95\columnwidth]{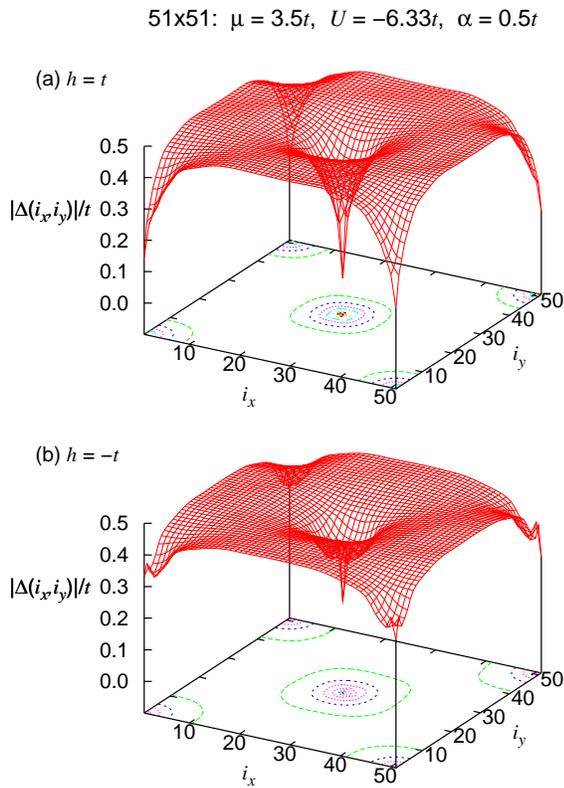}
  \caption{\label{fig:orderU-6.33} (Colour online) Order parameter of the vortex lattice in a pure 51$\times$51-site system for $\mu=3.5t$, $U=-6.33t$, and $\alpha=0.5t$; for (a) $h=t$ and (b) $h=-t$. Contour projection in steps of $0.05t$.
  }
\end{figure}

\begin{figure}
  \includegraphics[width=0.9\columnwidth]{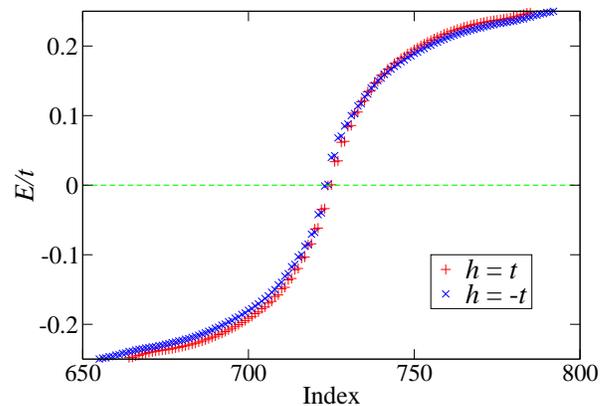}
  \caption{\label{fig:eigmu3.5_a0.5} (Colour online) Quasiparticle spectra for $h=t$ (red plus) and $h=-t$ (blue cross) in the pure 51$\times$51 vortex lattice for $\mu=3.5t$, $U=-6.33t$, and $\alpha=0.5t$. The index numbers the eigenvalues.
  }
\end{figure}

In this subsection, we illustrate that when SO coupling is relatively weak, vortex structure can be markedly different for the two opposite directions of the Zeeman field even in a pure vortex state. It has been found in Ref.~\onlinecite{Shitade2015} that the weaker the SO coupling, the closer the average angular momentum per Cooper pair to $-\hbar$ in the ($p_x-ip_y$)-dominated states. Thus, in terms of the direction of the Zeeman field (or alternatively the external field applied to create the vortex lattice) one can make one of the two intrinsic chiralities strongly manifest in the vortex core structure.

Shown in Fig.~\ref{fig:orderU-6.33} is the order parameter in the pure 51$\times$51 vortex lattice for $\mu=3.5t$, $\alpha=0.5t$, and $U=-6.33t$ ($\Delta\simeq 0.48t$, $E_0\simeq 0.21t$, and $\nu=1$); for (a) $h=t$ and (b) $h=-t$. Compared with the order parameter for $\mu=3.5t$ and $\alpha=t$ shown in Fig.~\ref{fig:orderU-5.25}(a) (for $h=-t$, but it is very similar to the order parameter for $h=t$), it can be seen in Fig.~\ref{fig:orderU-6.33}(a) that for $h=t$, the coherence length in the sense of a recovery length of the order parameter from the vortex centre to the bulk value is much larger for $\alpha=0.5t$ than $\alpha=t$. Note that the system size, namely, the size of the unit cell of the vortex lattice is 41$\times$41 and 51$\times$51, respectively, in Figs.~\ref{fig:orderU-5.25}(a) and \ref{fig:orderU-6.33}(a). In fact, we have tried the coupling constant of $U=-6.255t$ for $\alpha=0.5t$ that yields a similar bulk order parameter as for $\alpha=t$; however, the coherence (recovery) length in this case is so large that there is substantial overlap of vortex cores and it is not useful for comparison of the vortex structure with the $\alpha=t$ case.

Striking is the difference between the $\alpha=0.5t$ and $\alpha=t$ cases for $h=-t$ [Figs.~\ref{fig:orderU-6.33}(b) vs \ref{fig:orderU-5.25}(b)] and also between positive and negative $h$ in Figs.~\ref{fig:orderU-6.33}(a) and ~\ref{fig:orderU-6.33}(b). As can be seen in Fig.~\ref{fig:orderU-6.33}(b), for $h=-t$ as the order parameter tends toward zero at the vortex centre, it oscillates and is enhanced slightly around the vortex centre. At first glance, this is reminiscent of the $p_x+ip_y$ order parameter induced in the vortex core with vorticity $+1$ in the $p_x-ip_y$ domain of a chiral $p$-wave superconductor.\cite{Ichioka2002} In our calculation, however, the vorticity is $-1$ and thus parallel to the major chirality for $h=-t$. Moreover, we only have the superconducting order parameter stemming from spin-singlet $s$-wave pairing, and the enlargement of the vortex core for $h=t$ and the enhancement around the vortex centre for $h=-t$ are happening within the same $s$-wave order parameter. Both underlying chiralities $\pm 1$ are always present and mixed except for $\alpha \ll t$,\cite{Shitade2015} and it is not clear as to whether there is a way to define a unique order parameter for each of the two chiralities separately. It is apparent nonetheless that some extra order is induced in the $h=-t$ system inside the vortex core, increasing the superconducting order somewhat.

\begin{figure}
  \includegraphics[width=0.9\columnwidth]{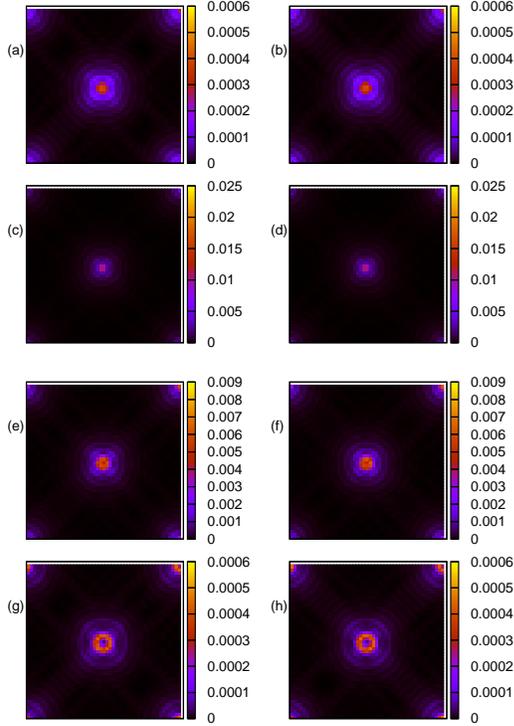}
  \vspace{.2cm}
  \caption{\label{fig:wfU-6.33_alpha0.5} (Colour online) Particle (left column) and hole (right column) probability amplitudes of the Majorana bound state for the systems shown in Figs.~\ref{fig:orderU-6.33} and \ref{fig:eigmu3.5_a0.5}; for spin up (a),(b) and spin down (c),(d) for $h=t$, and for spin up (e),(f) and spin down (g),(h) for $h=-t$.
  }
\end{figure}

\begin{figure}
  \includegraphics[width=0.9\columnwidth]{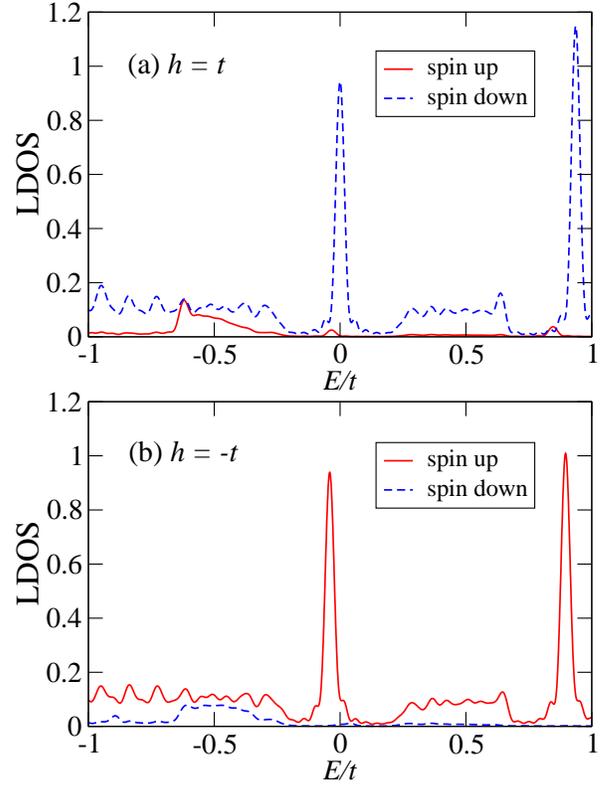}
  \vspace{.2cm}
  \caption{\label{fig:ldosU-6.33_alpha0.5} (Colour online) Spin-up (red solid curves) and spin-down (blue dashed curves) components of the LDOS as a function of quasiparticle energy at the vortex centre for the systems presented in Fig.~\ref{fig:wfU-6.33_alpha0.5}.   }
\end{figure}

The suppression and enhancement of the order parameter around the vortex centre for $h=t$ and $h=-t$, respectively, seen in Figs.~\ref{fig:orderU-6.33} (a) and \ref{fig:orderU-6.33}(b) are reflected in the quasiparticle spectra presented in Fig.~\ref{fig:eigmu3.5_a0.5}. One can see that the first few excited levels of the vortex bound states are higher for $h=-t$ than for $h=t$, with the minigap of $\sim 0.04t$ and $0.03t$, respectively ($\sim 0.05t$ for $\alpha=t$ for both $h=\pm t$). By comparison, the Majorana bound-state energy is not much different between $h=-t$ ($\sim 0.001t$) and $h=t$ ($\sim 0.0006t$). The ground-state energy of the system and equivalently the average energy gain per electron are also hardly different for $h=t$ and $h=-t$.
Bj\"ornson and Black-Schaffer have solved the BdG equations on the Hamiltonian (\ref{hamiltonian}) self-consistently (without the vector potential) for a vortex in a square lattice with open boundaries and have also found asymmetry in low-energy spectra between positive and negative $h$ for a given vorticity for $\alpha\approx 0.5t$.\cite{Bjornson2013}

The Majorana wavefunctions are shown in Fig.~\ref{fig:wfU-6.33_alpha0.5} for the above two systems; where the spin-up (a) particle and (b) hole probability amplitudes and the spin-down (c) particle and (d) hole probability amplitudes for $h=t$, and respective probability amplitudes for $h=-t$ in (e) and (f), and (g) and (h), are plotted as a function of $x$ and $y$ coordinates for the entire 51$\times$51 lattice. Once again, the particle (left column) and hole (right column) probability amplitudes are virtually identical in each spin component. For $h=t$, the bound-state peak of the spin-down component is strongly localized at the vortex centre [(c) and (d)], while the spin-up component, though much smaller in magnitude, extends substantially outside the core region [(a) and (b)].

Remarkably, the spin-down probability amplitudes are not peaked at the vortex centre for $h=-t$, as clearly visible in Figs.~\ref{fig:wfU-6.33_alpha0.5}(g) and \ref{fig:wfU-6.33_alpha0.5}(h). They are peaked at surrounding sites and though small in magnitude, have substantial extension outside the vortex core. The spin-up amplitudes are a little more confined: They are also suppressed right at the vortex centre, although it is not discernible in Figs.~\ref{fig:wfU-6.33_alpha0.5}(e) and \ref{fig:wfU-6.33_alpha0.5}(f). Furthermore, the spin-up amplitudes are one order of magnitude larger than the spin-down amplitudes, contrary to the $h=t$ case as well as the systems in Figs.~\ref{fig:wfU-8_h2.5} and \ref{fig:wfU-8_h-2.5} with strong SO coupling.
The LDOS at the vortex centre as a function of quasiparticle energy shown in Fig.~\ref{fig:ldosU-6.33_alpha0.5}(b) confirms the absence of the spin-down Majorana component at the vortex centre for $h=-t$, in stark contrast to the examples shown in Figs.~\ref{fig:ldosU-5.25_h1}, \ref{fig:ldosU-5.25_h-1}, \ref{fig:ldosU-8_h2.5} and \ref{fig:ldosU-8_h-2.5}, as well as the LDOS for $h=t$ in Fig.~\ref{fig:ldosU-6.33_alpha0.5}(a). The peak in the spin-up LDOS for $h=-t$ seen in Fig.~\ref{fig:ldosU-6.33_alpha0.5}(b) corresponds to the first excited state (the minigap at $\sim 0.04t$).

\begin{figure}
  \includegraphics[width=0.95\columnwidth]{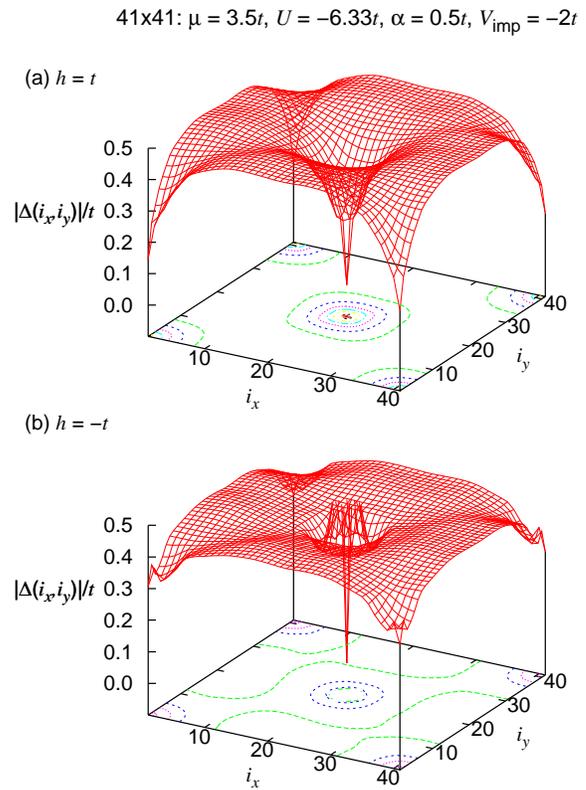}
  \caption{\label{fig:orderU-6.33_Vimp-2} (Colour online) Order parameter of the 41$\times$41 vortex lattice with $V_{\rm imp}=-2t$ at the centre site for $\mu=3.5t$, $U=-6.33t$, and $\alpha=0.5t$; for (a) $h=t$ and (b) $h=-t$. Contour projection in steps of $0.05t$.
  }
\end{figure}

Placing a nonmagnetic impurity at the vortex centre can enhance the unusual structure of the vortex core for $h=-t$. This is illustrated in Fig.~\ref{fig:orderU-6.33_Vimp-2}, where the order parameter is shown for the systems analogous to those presented in Fig.~\ref{fig:orderU-6.33}, but with $V_{\rm imp}=-2t$ at the centre of the 41$\times$41 vortex lattice, for (a) $h=t$ and (b) $h=-t$. For $h=t$, compared to the pure system (with 41$\times$41 lattice sites), the size of the vortex core is little changed by the nonmagnetic impurity, while its shape is modified slightly. For $h=-t$, on the other hand, the enhancement of the order parameter around the vortex centre is exaggerated by the presence of the impurity so much that the order parameter at its peaks is significantly larger than the bulk value.

\begin{figure}
  \includegraphics[width=0.95\columnwidth]{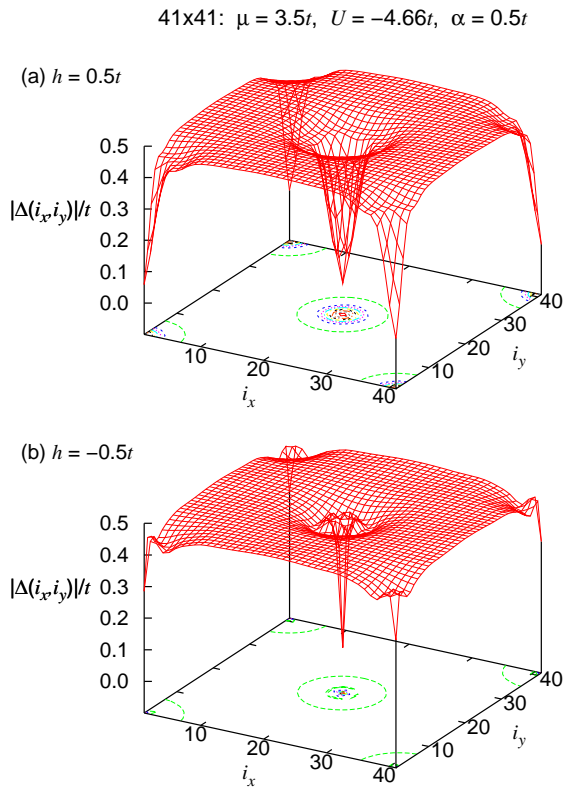}
  \caption{\label{fig:orderU-4.66_alpha0.5_h0.5} (Colour online) Order parameter of the vortex lattice in a pure 41$\times$41 system in trivial phase for $\mu=3.5t$, $U=-4.66t$, and $\alpha=0.5t$; for (a) $h=0.5t$ and (b) $h=-0.5t$. Contour projection in steps of $0.05t$.
  }
\end{figure}

Finally, we demonstrate the manifestation of chirality in the vortex lattice in the trivial phase. Shitade and Nagai\cite{Shitade2015} have found in the ($p_x-ip_y$)-dominated states that when $\alpha\ll t$, the average angular momentum per Cooper pair can be close to $-\hbar$ already in the trivial phase as $h$ ($>0$) approaches the critical value for the phase transition from the trivial to non-Abelian phase. We present in Fig.~\ref{fig:orderU-4.66_alpha0.5_h0.5} the order parameter in the pure 41$\times$41 vortex lattice for $\mu=3.5t$, $\alpha=0.5t$, and $U=-4.66t$, for (a) $h=0.5t$ and (b) $h=-0.5t$. These systems are in the trivial phase with $\nu=0$. The coupling constant $U$ has been chosen so as to make the bulk order parameter ($\Delta\simeq 0.48t$) similar to that in the non-Abelian systems for $\mu=3.5t$, $\alpha=0.5t$, and $h=\pm t$ discussed above. It can be seen in Figs.~\ref{fig:orderU-4.66_alpha0.5_h0.5}(a) and \ref{fig:orderU-4.66_alpha0.5_h0.5}(b) that the suppression and enhancement of the order parameter around the vortex centre for $h>0$ and $h<0$, respectively, are more substantial than in the non-Abelian systems shown in Figs.\ref{fig:orderU-6.33}(a) and \ref{fig:orderU-6.33}(b). This signifies the fact that the angular momentum carried by Cooper pairs governs vortex core structure, even though the current TSC model can be mapped onto a spinless $p$-wave superconductor with chiralities $\pm 1$ only in the nontrivial phase.\cite{Sato2010}

\subsection{\label{sec:minigap}Minigap}

\begin{figure}
  \includegraphics[width=0.9\columnwidth]{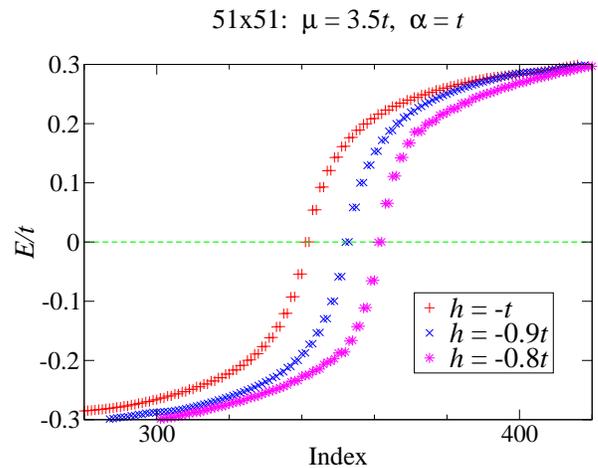}
  \caption{\label{fig:eigmu3.5_a1} (Colour online) Quasiparticle spectra for $\mu=3.5t$ and $\alpha=t$ in a pure 51$\times$51 vortex lattice. $U$ has been chosen so that $\Delta$ is roughly the same for different values of $h$. The index numbers the eigenvalues.
  }
\end{figure}

We show in Fig.~\ref{fig:eigmu3.5_a1} the quasiparticle spectra for $\mu=3.5t$ and $\alpha=t$ in the pure 51$\times$51 vortex lattice; for $h=-t$ (red plus), $h=-0.9t$ (blue cross), and $h=-0.8t$ (magenta star). The coupling constant has been chosen to be $U=-4.855t$ and $U=-5.05t$, respectively, for $h=-0.8t$ and $h=-0.9t$ so that the bulk order parameter $\Delta\approx 0.37t$ as for $h=-t$ ($U=-5.25t$). The bulk spectral gap (independent of the sign of $h$) is $E_0\simeq 0.18t$, $E_0\simeq 0.27t$, and $E_0\simeq 0.26t$ for $h=0.8t$, $h=0.9t$, and $h=t$, respectively. As mentioned in relation to Fig.~\ref{fig:eigmu-3} in Sec.~\ref{sec:chirality}, $\mu>0$ in these systems and the minigap is larger for $h<0$ than for $h>0$ by a very small amount (of the order of $10^{-3}t$). The minigap is $0.065t$, $0.058t$, and $0.054t$, respectively, for $h=-0.8t$, $h=-0.9t$, and $h=-t$. 

For these and other systems we have examined, we have not found any direct correlation between the minigap and the bulk order parameter nor the spectral gap. In this particular example (also for $h>0$), simply the smaller the $|h|$, the larger the minigap. The formula for the minigap for $s$-wave or chiral $p$-wave superconductors, $\sim \Delta/k_F\xi\sim \Delta E_0/k_F^2$, where $k_F$ is the magnitude of the Fermi wave vector and $\xi$ is the coherence length,\cite{deGennes,Tinkham} has been used in the literature for the $s$-wave TSC model with Rashba SO coupling and Zeeman field. For a given $\alpha$ and $h$, $E_0$ is determined uniquely from Eq.~(\ref{bulkspectrum}) once $\Delta$ is fixed. However, $k_F$ also depends on $\alpha$ and $h$. Thus, it is unclear as to whether this common formula applies to the minigap in $s$-wave TSC vortices with the two inherent chiralities. 

As discussed at the end of Sec.~\ref{sec:chirality}, the minigap is increased by a nonmagnetic impurity which the vortex is pinned by, and the increase in the minigap can be substantial as illustrated in Figs.~\ref{fig:ldosU-5.25_h-1} and \ref{fig:ldosU-8_h2.5}, depending on the major chirality and the strength of SO coupling. The larger the minigap, the shorter the minimum time required for braiding operation of vortices to avoid nonadiabatic transitions of the Majorana zero mode to excited states.\cite{Masaki2015} Therefore, further exploration of the effects of $\alpha$ and $h$ on the minigap, especially in the presence of a nonmagnetic impurity in the vortex core, is of great interest from the application point of view for topological quantum computation.

\section{\label{sec:conclusions}Conclusions}

In summary, we have performed a self-consistent study of the vortex lattice in a two-dimensional topological superconductor with an $s$-wave pairing interaction, Rashba spin-orbit coupling, and Zeeman field. When a vortex is pinned at a nonmagnetic impurity, one of the two intrinsic chiralities in the non-Abelian topological superconducting state can manifest itself in the vortex structure and affect the low-energy excitation spectrum for relatively weak spin-orbit coupling. One can make one chirality dominant over the other by changing the direction of the Zeeman field, or alternatively, the sign of the chemical potential.
In such states, (non-Majorana) vortex bound states are less influenced by the impurity if the dominant chirality is opposite to the vorticity, compared to the case where it is in the same direction as the vorticity. For stronger spin-orbit coupling, where orbital angular momentum is less of a ``good quantum number'', low-energy spectra tend to be more affected by the impurity regardless of the direction of the Zeeman field. The Majorana zero mode effectively remains a zero-energy bound state, and its spin is polarised according to the vorticity unless spin-orbit coupling is rather weak, in which case the spin polarization depends on the major chirality as well.

We have shown in Sec.~\ref{sec:weak-alpha} that when spin-orbit coupling is weak, the vortex core structure can be strikingly different for the two directions of the Zeeman field: The order parameter is suppressed and enhanced around the vortex centre when the major chirality is antiparallel and parallel, respectively, to the vorticity. We have demonstrated this phenomenon not only in non-Abelian topological phase, but also in trivial phase where the TKNN number is zero. The major chirality in such a trivial TSC state is the chirality of the dominant (or single) Fermi surface. The enhancement of the order parameter implies some extra superconducting order being induced. This occurs, however, in the vortex core region where the chirality and the vorticity are in the same direction, and appears to be counterintuitive in view of the vortex physics in spin-triplet $p_x\pm ip_y$ superconductors. Moreover, it is not immediately obvious how to map out chirality-based order parameters\cite{Bjornson2015,Shitade2015} from the sole order parameter in the model that is of $s$-wave pairing symmetry. The chiral nature of 2D $s$-wave TSC states and the role that the angular momentum carried by Cooper pairs plays in determining various properties of the system are to be explored in further detail in a future publication.

Finally, the suppression (the enlargement of the vortex core) and enhancement of the self-consistent order parameter around the vortex centre will be reflected in the supercurrent and hence the field distribution in the vortex core area. Thus, particularly in a 2D $s$-wave topological superconductor where spin-orbit coupling is relatively weak, manifestation of different chiralities can be detected by probing the field distribution in the vortex lattice, e.g., by NMR and by switching the direction of the applied magnetic field.

\section{\label{sec:acknowledgements}Acknowledgements}

We thank S. L. Goertzen and T. Kawakami for helpful discussions. K.T. is grateful to CCSE at Japan Atomic Energy Agency for hospitality, where part of the research was performed. This work was enabled in part by support provided by WestGrid (www.westgrid.ca) and Compute Canada Calcul Canada (www.computecanada.ca). Part of the calculation was also performed on the supercomputing system SGI ICE X at the Japan Atomic Energy Agency. The research was supported by the Natural Sciences and Engineering Research Council of Canada, and partially by JSPS KAKENHI Grant No. 26800197 and the ``Topological Materials Science'' (No. 16H00995) KAKENHI on Innovative Areas from JSPS of Japan.


\begin{thebibliography}{99}
\bibitem{Sato2009}
M. Sato, Y. Takahashi, and S. Fujimoto, Phys. Rev. Lett. {\bf 103}, 020401 (2009).
\bibitem{Sato2010}
M. Sato, Y. Takahashi, and S. Fujimoto, Phys. Rev. B {\bf 82}, 134521 (2010).
\bibitem{Sau2010}
J. D. Sau, R. M. Lutchyn, S. Tewari, and S. Das Sarma, Phys. Rev. Lett. {\bf 104}, 040502 (2010).
\bibitem{Alicea2010}
J. Alicea, Phys. Rev. B {\bf 81}, 125318 (2010).
\bibitem{Alicea2012}
J. Alicea, Rep. Prog. Phys. {\bf 75}, 076501 (2012).
\bibitem{Tewari2010}
S. Tewari, J. D. Sau, S. Das Sarma, Ann. Phys. {\bf 325}, 219 (2010).
\bibitem{Kopnin1991}
N. B. Kopnin and M. M. Salomaa, Phys. Rev. B {\bf 44}, 9667 (1991).
\bibitem{Read2000}
N. Read and D. Green, Phys. Rev. B {\bf 61}, 10267 (2000). 
\bibitem{Ivanov2001}
D. A. Ivanov, Phys. Rev. Lett. {\bf 86}, 268 (2001).
\bibitem{Stern2004}
A. Stern, F. von Oppen, and E. Mariani, Phys. Rev. B {\bf 70}, 205338 (2004). 
\bibitem{Stone2006}
M. Stone and S.-B. Chung, Phys. Rev. B {\bf 73}, 014505 (2006).
\bibitem{Fujimoto2008}
S. Fujimoto, Phys. Rev. B {\bf 77}, 220501(R) (2008).
\bibitem{Sato2009p-wave}
M. Sato and S. Fujimoto, Phys. Rev. B {\bf 79}, 094504 (2009).
\bibitem{Zhang2008}
C. Zhang, S. Tewari, R. M. Lutchyn, and S. Das Sarma, Phys. Rev. Lett. {\bf 101}, 160401 (2008).
\bibitem{Shitade2015}
A. Shitade and Y. Nagai, Phys. Rev. B {\bf 92}, 024502 (2015).
\bibitem{Frigeri2006}
P. A. Frigeri, D. F. Agterberg, I. Milat, and M. Sigrist, Eur. Phys. J. B {\bf 54}, 435 (2006).
\bibitem{Takigawa2001}
M. Takigawa, M. Ichioka, K. Machida, M. Sigrist, Phys. Rev. B {\bf 65}, 014508 (2001); see also references therein.
\bibitem{Ichioka2002}
M. Ichioka and K. Machida, Phys. Rev. B {\bf 65}, 224517 (2002).
\bibitem{Anderson1959}
P. W. Anderson, J. Phys. Chem. Solids {\bf 11}, 26 (1959).
\bibitem{Kato2000}
Y. Kato, J. Phys. Soc. Jpn. {\bf 69}, 3378 (2000).
\bibitem{Kato2002}
Y. Kato and N. Hayashi, J. Phys. Soc. Jpn. {\bf 71}, 1721 (2002).
\bibitem{Masaki2014}
Y. Masaki, Y. Kato,  J. Phys.: Conf. Ser. {\bf 568}, 022028 (2014).
\bibitem{Masaki2015}
Y. Masaki and Y. Kato, J. Phys. Soc. Jpn. {\bf 84}, 094701 (2015).
\bibitem{Volovik1999}
G. E. Volovik, JETP Lett. {\bf 70}, 609 (1999).
\bibitem{Ioselevich2012}
P. A. Ioselevich, P. M . Ostrovsky, and M. V. Feigel’man, Phys. Rev. B {\bf 86}, 035441 (2012). 
\bibitem{Schnyder2008}
A. P. Schnyder, S. Ryu, A. Furusaki, and A. W. W. Ludwig, Phys. Rev. B {\bf 78}, 195125 (2008). 
\bibitem{deGennes}
P. G. de Gennes, \emph{Superconductivity of Metals and Alloys} (Westview Press, Boulder, 1999).
\bibitem{Matetskiy2015}
A. V. Matetskiy, S. Ichinokura, L. V. Bondarenko, A. Y. Tupchaya, D. V. Gruznev, A. V. Zotov, A. A. Saranin, R. Hobara, A. Takayama, and S. Hasegawa, Phys. Rev. Lett. {\bf 115}, 147003 (2015).
\bibitem{Li2016}
L. J. Li, E. C. T. O'Farrell, K. P. Loh, G. Eda, B. \"Ozyilmaz, and A. H. Castro Neto, Nature (London) {\bf 529}, 185 (2016).
\bibitem{Nagai2016}
Y. Nagai, S. Hoshino, and Y. Ota, Phys. Rev. B {\bf 93}, 220505 (2016).
\bibitem{Covaci2010}
L. Covaci, F. M. Peeters, and M. Berciu, Phys. Rev. Lett. {\bf 105}, 167006 (2010).
\bibitem{Nagai2012}
Y. Nagai, Y. Ota, and M. Machida, J. Phys. Soc. Jpn. {\bf 81}, 024710 (2012).
\bibitem{Sakurai2003}
T. Sakurai and H. Sugiura, J. Comput. Appl. Math. {\bf 159}, 119 (2003).
\bibitem{Nagai2013}
Y. Nagai, Y. Shinohara, Y. Futamura, Y. Ota, and T. Sakurai, J. Phys. Soc. Jpn. {\bf 82}, 094701 (2013).
\bibitem{Nagai2015}
Y. Nagai, Y. Ota, and M. Machida, J. Phys. Soc. Jpn. {\bf 84}, 034711 (2015).
\bibitem{TKNN1982}
D. J. Thouless, M. Kohmoto, M. P. Nightingale, and M. den Nijs, Phys. Rev. Lett. {\bf 49}, 405 (1982).
\bibitem{Ishikawa1987}
K. Ishikawa and T. Matsuyama, Nucl. Phys. B {\bf 280}, 523 (1987).
\bibitem{Volovik2009}
G. E. Volovik, \emph{The Universe in a Helium Droplet} (Oxford University Press, Oxford, 2009).
\bibitem{Gurarie2011}
V. Gurarie, Phys. Rev. B {\bf 83}, 085426 (2011). 
\bibitem{Nagai2014oval}
Y. Nagai, Y. Ota, and M. Machida, J. Phys. Soc. Jpn. {\bf 83}, 094722 (2014).
\bibitem{Stone2004}
M. Stone and R. Roy, Phys. Rev. B {\bf 69}, 184511 (2004).
\bibitem{Takigawa2000}
M. Takigawa, M. Ichioka and K. Machida, J. Phys. Soc. Jpn. {\bf 69}, 3943 (2000).
\bibitem{Kawakami2016}
T. Kawakami, private communication.
\bibitem{Wu2012}
L.-H. Wu, Q.-F. Liang, Z. Wang and X. Hu, J. Phys.: Conf. Ser. {\bf 393}, 012018 (2012).
\bibitem{Weisse2006}
A. Wei{\ss}e, G. Wellein, A. Alvermann and H. Fehske, Rev. Mod. Phys. {\bf 78}, 275 (2006).
\bibitem{Nagai2014top}
Y. Nagai, H. Nakamura, and M. Machida, J. Phys. Soc. Jpn. {\bf 83}, 064703 (2014).
\bibitem{Hu2013}
H. Hu, L. Jiang, H. Pu, Y. Chen, and X.-J. Liu, Phys. Rev. Lett. {\bf 110}, 020401 (2013).
\bibitem{Goertzen2016}
S. L. Goertzen, Y. Nagai, and K. Tanaka (unpublished).
\bibitem{Mao2010}
L. Mao and C. Zhang, Phys. Rev. B {\bf 82}, 174506 (2010).
\bibitem{Bjornson2013}
K. Bj\"ornson and A. M. Black-Schaffer, Phys. Rev. B {\bf 88}, 024501 (2013).
\bibitem{Tinkham}
M. Tinkham, \emph{Introduction to Superconductivity} (McGraw-Hill, New York, 1996).
\bibitem{Bjornson2015}
K. Bj\"ornson and A. M. Black-Schaffer, Phys. Rev. B {\bf 91}, 214514 (2015).
\end{thebibliography}

\end{document}